\documentclass[11pt]{article}
\usepackage[utf8]{inputenc}
\usepackage[margin=1in]{geometry}
\usepackage{bm}
\usepackage{nameref}
\usepackage{amsmath}
\usepackage{authblk}
\usepackage{natbib}
\usepackage{amssymb}
\usepackage{tikz}
\usepackage{multirow}
\usepackage{multicol}
\usepackage{url}
\usepackage{array}
\usepackage{subcaption}
\usetikzlibrary{shapes,arrows,calc}

\title{Preparing urban mobility for the future of work}

\author[1]{Nicholas S. Caros}
\author[2,*]{Jinhua Zhao}

\affil[1]{Department of Civil and Environmental Engineering, Massachusetts Institute of Technology, Cambridge, MA 02139, USA}
\affil[2]{Department of Urban Studies and Planning, Massachusetts Institute of Technology, Cambridge, MA 02139, USA}
\affil[*]{Corresponding Author: jinhua@mit.edu}

\begin{document}

\maketitle

\section{Abstract}

A gradual growth in flexible work over many decades has been suddenly and dramatically accelerated by the COVID-19 pandemic.
The share of flexible work days in the United States is forecasted to grow from 4\% in 2018 to over 26\% by 2022.
This rapid and unexpected shift in the nature of work will have a profound effect on the demand for, and supply of, urban transportation. 
Understanding how people make decisions around where and with whom to work will be critical for predicting future travel patterns and designing mobility systems to serve flexible commuters.
To that end, this paper establishes a formal taxonomy for describing possible flexible work arrangements, the stakeholders involved and the relationships between them. 
An analytical framework is then developed for adapting existing transportation models to incorporate the unique dynamics of flexible work location choice. 
Several examples are provided to demonstrate how the new taxonomy and analytical framework can be applied across a broad set of scenarios.
Finally, a critical research agenda is proposed to create both the empirical knowledge and methodological tools to prepare urban mobility for the future of work.

\hfill\break
\noindent\textit{Keywords}: Future of work, flexible work, sustainable mobility, transportation modeling
\newpage

\section*{Introduction}

Remote and flexible work have been gaining popularity for decades, but until 2020 remained a relatively small share of total employment.
Throughout that time, the academic community expressed considerable interest in the impact of flexible work on transportation, but development of analytical tools focused justifiably on understanding, modeling, predicting and improving travel for the dominant group:  traditional commuters. 
The absence of comprehensive transportation models for flexible work is no longer justifiable, however; the COVID-19 pandemic has dramatically accelerated previous trends, forcing many organizations and their staff to become accustomed to flexible work and inspiring the development of better digital collaboration and productivity tools. 
This new familiarity, along with the rise of advanced communication technology has, in turn, led to dramatic and likely enduring changes in flexible work adoption. 
Hundreds of top employers have transitioned a significant number of their employees to fully remote positions, and it is expected that a quarter of all work days will continue to take place remotely after the acute public health threat of the COVID-19 pandemic subsides \citep{barrero2021working}.

A large share of flexible work will have tremendous social impacts in the long term, creating challenges for a number of stakeholders. 
Urban transportation systems, designed largely to accommodate the rush of commuters for a few hours in the morning and evening, will need to adjust to serve a more evenly distributed demand profile. 
Employers will be obliged to develop smart flexible work policies and real estate portfolios that balance employee preferences, productivity, budget and culture.
Commercial landlords and co-working providers will be forced to adapt their offerings, locations and business models to cater to organizations who need flexibility and convenience. 
Retail stores and food service businesses dependent on office workers or commuters may suffer as a result of less in-person work, but those based in residential neighborhoods could flourish. 
Finally, policy makers and planners responsible for land use and infrastructure will need to rethink zoning and service provision to promote thriving and vibrant communities that meet the needs of flexible workers. 

The new cohort of flexible workers, representing a significant portion of the global economy, is likely to have different motivations than the ``teleworkers'' of the 1990s and 2000s. 
Understanding and adapting to their travel patterns and mode choices will require revisiting past assumptions and empirical results.
It is doubtful that the impacts of a fourfold increase in flexible work can simply be extrapolated from previous trends; the scale of the societal transition is more akin to a \emph{phase shift}, one that could give rise to an entirely new ecosystem of services and organizations.
New analytical tools and a vocabulary that can capture the new relationships, business models, and policies that might arise in response are urgently needed. 

The shift towards flexible work is not simply a series of challenges to overcome, however. 
We are experiencing a liminal moment in which we can re-examine the status quo and begin to shape a more sustainable, equitable future. 
As \citeauthor{Beck_Hensher_2021} put it in a recent article on the potential impacts of flexible work, ``The COVID-19 pandemic provides an opportunity to allow decision-makers to take a hard look at the assumptions that underlie many of the decisions made on transport and land use futures'' \citep{Beck_Hensher_2021}.
Transit agencies could shift excess peak resources to extend their service areas and compete for a wider range of trips, including trips by those who are not traditional commuters.
Employers could allow their employees the flexibility to work from home (or closer to home), reducing commutes and thus transportation-related pollution and stress.
Alternatively, introducing satellite offices in co-working spaces could shorten commutes while retaining in-person collaboration and promoting cross-organizational learning, as well as mitigate feelings of social isolation. 
Planners could create lively neighborhoods throughout an urban area by allowing for more residential land use in newly vacant downtown offices and more commercial land use for flexible work in neighborhood centers. 

All of the stakeholder decisions to meet the challenges and seize the opportunities presented by flexible work will directly affect, or be affected by, urban mobility. 
A broad, multi-disciplinary base of theoretical and empirical research, centered around urban mobility and the future of work, is therefore critical in order to enable evidence-based decision making.
This article is intended to lay the foundation for that research by creating a versatile framework within which research can be classified, and a taxonomy for describing the relationships between mobility and flexible work. 
Furthermore, it is intended to be a call to action for the academic community, identifying research gaps and opportunities through a thorough examination of existing literature across a variety of fields.
While a small number of recent papers have hypothesized about the long-term impacts of flexible work on mobility, this is the first paper to offer a systematic, interdisciplinary literature review and a comprehensive framework and taxonomy for future research.  

Before continuing, some remarks on the terminology used in this paper and its overall scope. First, ``flexible work'' is used to represent an arrangement wherein an employee is given some degree of choice over their work location. 
This is either more general or refers to something markedly different than other similar terms, including ``teleworking'', ``remote work'', ``work from home'' or ``work from anywhere''.
Second, this paper is primarily interested in the long-term effects of flexible work on urban mobility, and as such it does not include a discussion of the temporary changes to everyday mobility patterns that occurred during the height of the COVID-19 pandemic and associated economic restrictions. 
There is already a wide body of published research on that topic; the reader is referred to \citet{borkowski2021lockdowned} for an excellent summary.

The remainder of this paper is structured as follows. 
Section~\ref{sec:trends} summarizes the decades-long rise of flexible work along with recent surveys that forecast future trends.
Section~\ref{sec:lit_review} reviews existing literature in the area of flexible work across several relevant fields of research. 
Section~\ref{sec:taxonomy} then enumerates possible flexible work arrangements, the stakeholders involved, and the relationships between them, thus developing a taxonomy and a classification system.
Section~\ref{sec:framework} applies this taxonomy to establish a structured framework for describing and categorizing research concerned with various aspects of flexible work.
Section~\ref{sec:research} identifies a set of critical research gaps to be addressed.
Finally, Section~\ref{sec:conclusion} concludes the paper with a summary and further discussion. 

\section{Flexible work trends} \label{sec:trends}

To understand how flexible work might evolve in the future, it is important to review historical trends and causal factors. 
Flexible work has been proposed at least as far back as the 1950s \citep{wiener1950speech}, and became sufficiently popular by the 1980s to become the subject of several economic and organizational behavior studies \citep{Olson_1983, shamir1985work}. 
Surveys of flexible workers conducted around this time implies some participation, but the phenomenon remained fairly uncommon and difficult to measure \citep{Mokhtarian_Salomon_Choo_2005}.
A number of societal trends contributed to the growth of flexible work thereafter. 
Advances in digital and communication technology enabled the first truly flexible workers; personal computers, the internet, email and eventually video conferencing have progressively improved remote collaboration \citep{DalFiore_Mokhtarian_Salomon_Singer_2014}. 

The number of people working at least one day per month at home in the United States rose from 4 million in 1990 to 23.6 million in 2001 \citep{pratt2003telework} as technology continued to improve. 
New societal factors, such as increased globalization, the rise of the ``gig economy'', and online sales platforms have made it easier to substitute freelancing and self-employment for full-time salary work \citep{smith2016gig}.
Yet even in 2018, flexible work accounted for only 5\% of all worked days in the United States \citep{barrero2021working}. 

Then, in early 2020, the COVID-19 pandemic forced most workplaces to close due to the risk of spreading the virus. 
By May 2020 over 60\% of worked days in the U.S. were taking place at home \citep{barrero2021working}, a twelvefold increase compared to the pre-pandemic norm.
Throughout the pandemic, employers began to rethink their long-term flexible work policies.
Much like the inertia that kept traditional in-person work the dominant arrangement long after technology was sufficient to permit remote working, the inertia of pandemic-related remote work has begun to increase the desire for permanent full-time remote work. 
Many large employers, including Salesforce, Facebook and Google, have announced flexible work policies that include full-time remote options \citep{cutterWSJ}.
 
\begin{figure*}[ht!]
    \begin{center}
        \frame{\includegraphics[width=0.9\textwidth]{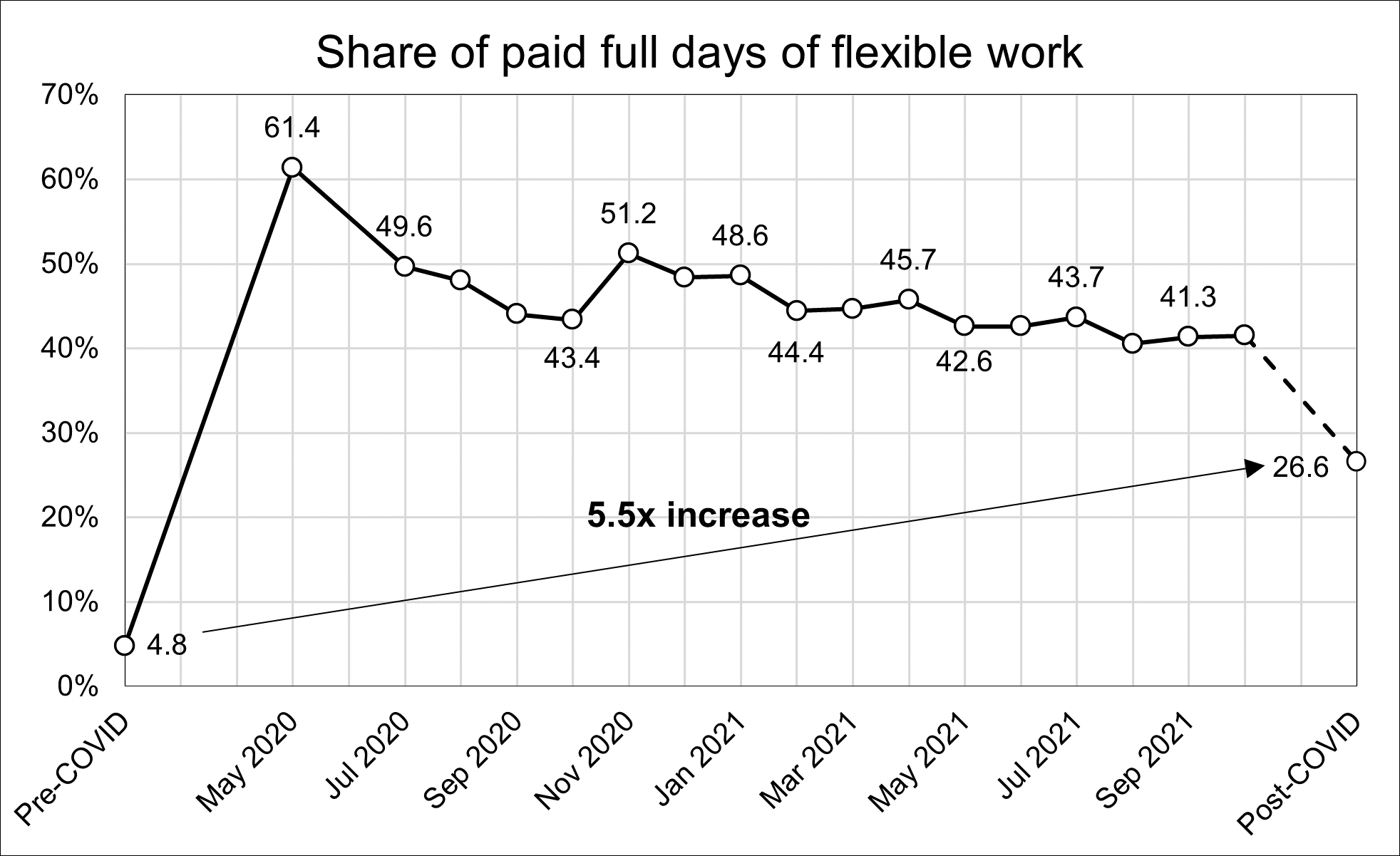}}
        \caption{Flexible work as a percentage of all worked days before, during, and after COVID-19. Adapted from \citet{barrero2021working}. Note: Pre-COVID share based on 2017-2018 American Time Use Survey; post-COVID share is the share anticipated by employers.}
        \label{fig:fw_trends}    
    \end{center}
\end{figure*}

The degree to which flexible work will remain after the public health threat of COVID-19 subsides has been the subject of a lot of speculation and survey efforts.
The surveys vary in methodology and timing, making the results difficult to compare.
\citet{barrero2021working} conducted a comprehensive longitudinal survey of 22,500 working-age Americans between May 2020 and November 2020.
The survey found that employers intend for 26.6\% of all worked days to take place remotely going forward (see Figure~\ref{fig:fw_trends}), while the average worker would prefer flexible work 47\% of the time.
A PwC survey of 1,200 U.S. workers around the same time found that, on average, workers would prefer 56\% of their work days to be flexible after COVID-19 has passed \citep{pwc2021}.
The employer intention results from \cite{barrero2021working} are similar to a survey of HR professionals and hiring managers conducted in April 2020 \citep{ozimek2020future}.
The results are heterogeneous across job sectors, income levels and gender, suggesting that the benefits and impacts of flexible work will not be shared evenly. 
Surveys have also found a racial difference in the desire to return to the office full time \citep{economist2021}.

Surveys issued in 2021 have often found a greater preference for flexible work than those issued early in the pandemic.
A Harvard Business School survey of 1,500 U.S. professionals in March 2021 \citep{hbs2021} found that 88\% would prefer at least 2 days per week of flexible work. 
In their longitudinal survey from May 2020 to March 2021, \citet{barrero2021working} has found that the number of people wanting only flexible work has increased over time. 
Preferences for flexible work may be growing as a return to the office becomes more realistic and workers begin to consider the downsides (commuting, etc.) more closely. 
People may also have become more comfortable with flexible work and virtual communication over time.
Finally, it may be that people began to enjoy the additional location choices provided by flexible work as pandemic-related restrictions on travel, retail outlets and social gatherings were lifted.

Employees in certain sectors of the economy, typically service and knowledge work sectors, appear to value flexible work more than the average worker.
Surveys of staff at higher education institutions have found a significant preference for flexible work; in December 2020 Duke University found that staff would prefer flexible work for 70\% of working days, with 91\% of respondents preferring more than 1 day of flexible work \citep{duke2021}. 
Boston University \citep{bu2021} and the University of Michigan \citep{michigan2021} found similar trends for their workforces. 
Technology company executives surveyed in September 2020 indicated that, on average, 34\% of their employees will have entirely flexible work after the pandemic \citep{chavez2020}. 

Global surveys have been less common than US-based surveys, but there have been several notable examples.
Boston Consulting Group collected responses from 209,000 people across 190 countries in October and November of 2020 \citep{bcg2020}. 
The study finds that, like many of the US studies, approximately 9 in 10 workers would prefer flexible work at least 2 days per week.
The higher preference for flexible work among knowledge and office workers is also observed globally.
Interestingly, the study finds that workers in less developed countries are most interested in flexible work and hypothesize that this could be due to differences in relative transportation costs.
A December 2020 - January 2021 international survey by McKinsey found that, on average, corporate and government workers preferred 52\% of their days to be flexible.
Workers in Latin America and Australia favored flexible work the most, while workers in Asia and Europe were least interested \citep{mckinsey2021}.
A survey of approximately 200 workers in India found that the share of people with at least one day of flexible work is expected to double as a result of the pandemic, from 34\% to 76\% \citep{Nayak_Pandit_2021}.  

\section{Literature review} \label{sec:lit_review}

This section summarizes the state of the art from four disciplines as they relate to flexible work: travel behavior, transportation science, land use and real estate, and organizational behavior. 
Section~\ref{sec:travel_behavior} (Travel Behavior) includes the micro-level decision making processes and associated psychological factors that contribute to travel choices regarding flexible work.
Section~\ref{sec:transportation_science} (Transportation Science) focuses on measuring aggregate impacts of flexible work on travel demand, and analytical models of transportation systems that explicitly incorporate flexible work travel behavior. 
There has been considerable speculation about the effects of flexible work on cities; these studies are summarized in Section~\ref{sec:land_use} (Land Use and Real Estate).
Finally, Section~\ref{sec:org_beh} (Organizational Behavior) includes research related to individual and organizational attitudes towards flexible work and how such arrangements affect organizational outcomes. 

\subsection{Travel behavior} \label{sec:travel_behavior}

Flexible work was a popular research topic in the 1990s and several discrete choice models were estimated based on survey responses. 
\citet{bagley1997analyzing} and \citet{stanek1998developing} conducted surveys of workers in California to elicit preferences for working from home and from a flexible work center.
\citet{Mokhtarian_Salomon_1997} found that attitudes towards work, family and commuting are more important than socio-demographic factors in determining preferences towards flexible work.
This is similar to the results from \citet{Vana_Bhat_Mokhtarian_2008}.
In addition to estimating a discrete choice model, \citet{yeraguntla2005classification} offers a taxonomy of flexible work arrangements.
Given that telecommunication technology was not very mobile at the time, the authors only considered two flexible work locations: home and a regional teleworking center. 

\citet{Pouri_Bhat_2003} includes several occupational factors in a flexible work choice model, finding that part time workers and employees of private companies are more likely to choose flexible work, while those requiring daily face-to-face interactions are less likely to choose flexible work. 
\citet{Sener_Bhat_2011} also included work characteristics in estimating a copula-based sample selection model using household travel survey data from Chicago. 
\citet{Tang_Mokhtarian_Handy_2011} and \citet{Singh_Paleti_Jenkins_Bhat_2013} review the impact of the built environment on the propensity to work from home, and confirms the existence of several nuanced effects, such as greater perceived regional accessibility leading to less frequent flexible work. 
\citet{arabikhan2017telecommuting} demonstrates that a fuzzy rule based network, rather than the traditional discrete choice framework, can improve modeling of flexible work adoption.

While these discrete choices studies were being conducted, there was also an effort to develop a behavioral theory to explain the rationale behind the observed decisions.
In a remarkable series of papers, \citeauthor{Mokhtarian_Salomon_1994} create a conceptual model to explain the desire for flexible work in terms of constraints, attitudes, personal satisfaction and utility \citep{Mokhtarian_Salomon_1994, Mokhtarian_Salomon_1996a, Mokhtarian_Salomon_1996b}.
\citet{vanWee_Witlox_2021} point out that many traditional travel behavior concepts (utility theory, social practice theory, time geography, and theory of planned behavior) would predict a significant increase in flexible work after COVID-19 due to increased familiarity with virtual communication tools and lasting changes in social norms.
The authors recommend research into policy responses for long-term changes in activity and travel behavior as a result of the COVID-19 pandemic.

Much of the effort in modeling flexible work decisions has focused on the frequency and duration of flexible work, rather than the location and the choice to co-locate with others \citep{Handy_Mokhtarian_1996b, mokhtarian1998synthetic}.
This is partly due to implicit assumptions that flexible workers are making a binary choice: work at an office or work from home. 
Even recent comprehensive frameworks that include duration of flexible work do not consider location choice or the impact of personal relationships \citep{asgari2014choice, Asgari_Jin_2015, paleti2016generalized}. 
While anecdotal, recent interviews with flexible workers in the New York City area found that many enjoy working from non-home, non-work locations for increased productivity and "a change of scenery" \citep{stiles2019working}. 
This suggests a latent demand for alternative work locations that may become more prevalent as the COVID-19 related restrictions on travel and group activities begin to ease. 

\subsection{Transportation science} \label{sec:transportation_science}

Even as flexible work (known at the time as ``telework'') was in its infancy, \citet{harkness1977selected} hypothesized that adoption of flexible work could have significant impacts on transportation systems. 
Clearly flexible work has long been of interest to transportation researchers, but most of the literature is focused on empirical studies rather than analytical models that connect flexible work and transportation.

A seminal report by \citet{mokhtarian1991empirical} and review by \citet{Nilles_1988} provide a good summary of early empirical research. 
\citeauthor{mokhtarian2002telecommunications} is a pioneer in this field of study, publishing dozens of influential papers over several decades \citep{mokhtarian1990typology, mokhtarian1991telecommuting, mokhtarian1995methodological, Handy_Mokhtarian_1995, mokhtarian1997modeling, mokhtarian1998synthetic,  mokhtarian2002telecommunications}.
From the very beginning they identified the need for flexible work to be incorporated into travel demand models \citep{Handy_Mokhtarian_1996a}. 
More recent empirical research includes studies of how flexible work has affected road congestion in Australia \citep{cleary2010more}, Norway \citep{julsrud2012business}, Iran \citep{vakilian2019modeling}, and Sweden \citep{bieser2021impacts}. 

In addition to the aggregate impact on travel, researchers have also investigated differences in travel patterns between flexible and traditional workers.
Interestingly, \citet{hong2002travel} finds that home-based flexible workers are less likely to travel during the morning rush hour, but do often travel during the evening rush hour.
A more recent paper found similar trends for knowledge workers in the United States \citep{Stiles_Smart_2020}.
This suggests that flexible workers are generally sticking to typical work schedules, and are eager to pursue non-work activities after they have finished work.
These results could have a significant impact for transportation providers, as the effect of flexible work of peak demand may not be symmetric. 
\citet{Handy_Mokhtarian_1995} notes that the popularity of flexible work and its effect on travel demand is related to many other travel demand management policies, including congestion pricing, office parking subsidies and zoning. 

The impact of flexible work on non-work travel behavior has long been debated.
Early research finds that the average number of trips on flexible work days increases, but the distance travelled decreases \citep{Mokhtarian_Varma_1998}. 
A PhD thesis analyzing U.S. travel behavior data from 1995 confirms these results, and finds that flexible workers and their families have a greater average annual travel distance \citep{hong2002travel}.
Recent papers have found more daily travel overall for flexible workers in the United States  \citep{deAbreuSilva_Melo_2018, zhu2018metropolitan,Su_McBride_Goulias_2021}.
On the other hand, \citet{choo2005does} finds that flexible work has little-to-no impact on overall vehicle-miles travelled in the United States; \citet{deAbreuSilva_Melo_2018b} encounters a similar result for single-household workers in the United Kingdom. 
\citet{Kim_Choo_Mokhtarian_2015} finds that non-work travel is higher for Korean families whose primary breadwinner is able to work remotely, but only if there are fewer vehicles than employed adults in the household. 

\citet{deAbreuSilva_Melo_2018} finds that part-time flexible workers have a longer average commute and are more likely to commute by private car, leading to more travel overall and less sustainable travel choices. 
\citet{Ory_Mokhtarian_2006}, however, suggests that the decision to move further from work tends to precede the start of any part-time flexible work, while those who move after beginning part-time flexible work actually move closer to their workplace.
\citet{Salomon_Mokhtarian_2008} argues that the differences in results are often due to heterogeneity among flexible workers, as well as a lack of consistency in defining flexible work and in survey methodology.

The opposite effect, whether discretionary trips affect the decision to choose flexible work, has also been investigated \citep{Asgari_Jin_2017}. 
The authors find that people with the option for flexible work are more likely to do so on days where they plan a discretionary activity, and the length of the discretionary activity is predictive of the decision to engage in a full day of flexible work.

Despite the evidence of different travel patterns, relatively few models that incorporate the effects of flexible work on urban transportation systems have been developed. 
Two seminal papers in modeling the transportation impacts of flexible work, \citet{Nagurney_Dong_Mokhtarian_2002, Nagurney_Dong_Mokhtarian_2003}, provide an equilibrium traffic flow model formulation and solution method that includes the option of teleworking at multiple alternative destinations.
The authors introduce virtual links to the network to represent the teleworking alternatives.
\citet{Pawlak_Polak_Sivakumar_2015} also allows for teleworking in their comprehensive econometric model of the joint choice of activity, duration, mode and route. 
Similarly, \citet{de2007substitution} incorporates the preference for working at home or out of home directly into the utility function in order to estimate the trade-off between the two working arrangements. 

One paper was found that includes simulation of a transportation system with flexible work locations \citep{Ge_Polhill_Craig_2018}.
The authors use an agent-based regional travel demand model to evaluate the effect of flexible workplaces on commuting distances.
Interestingly, they find that requiring co-location of teams can lead to a worse outcomes than the status quo under certain conditions. 
The study does not include any mathematical modeling or productivity considerations, however. 

Finally, there have been some efforts among transportation researchers to develop analytical models for the \emph{long-term} impacts of the COVID-19 pandemic on transportation demand.
\citet{zhang2021long} uses a urban equilibrium model to predict the effect of various post-pandemic policies on long term mobility in China.
The authors find that while flexible work produces more sustainable mobility outcomes, these effects are largely negated if shared mobility (public transit and carsharing) become less popular. 
\citet{habib2021examining} use an integrated land use model to estimate that car ownership and commuting distance could be expected to increase as a result of the pandemic.

\subsection{Land use and real estate} \label{sec:land_use}

Transportation and land use are intrinsically linked, so it is important to consider how changes in mobility patterns due to flexible work can impact the built environment, and vice versa. 
The US Department of Transportation has been interested in flexible work as a travel demand management strategy for some time; in the 1990s, the department funded the establishment of Residential Area-Based Offices (RABO), a precursor to co-working spaces, with the intention of reducing the air pollution associated with commuting \citep{Bagley_Mannering_Mokhtarian_1994}.
\citeauthor{mokhtarian1998synthetic} has been arguing for land-use policies that encourage co-working centers in residential and mixed-use developments since the 1990s in order to reduce commutes \citep{mokhtarian1998synthetic}. 

\citet{Saxena_Mokhtarian_1997}, in an early pilot study, found that flexible workers visit destinations that are both closer to home and more even distributed geographically on days that they work from home compared to days that they commute to an office.
These results were confirmed by \citet{Asgari_Jin_Rojas_2019} for the New York City area.
This suggests a re-alignment of demand away from commuting corridors and commercial districts towards residential areas and neighborhood centers.
A recent paper also found similar results using the National Household Travel Survey, noting that flexible workers have more complex and varied schedules, and that they visit more locations than regular commuters \citep{Su_McBride_Goulias_2021}.
The authors note that flexible workers remain at home on only 20\% of work days and often chose to work in a location outside the home. 

Another related area of research is modeling job accessibility in a partial flexible environment.
\citet{muhammad2008modelling} introduces virtual spaces into the accessibility modeling framework and find that flexible work increases job accessibility overall, with greater benefits in rural areas.
Several subsequent papers by many of the same authors explore this area in greater detail \citep{van2012ict,van2013information, bartosiewicz2015use}.

Theoretical models of urban economies generally suggest that increased flexible work will lead to decentralization, with areas located at a medium distance from urban centers experiencing the greatest increase in demand \citep{Lund_Mokhtarian_1994,muhammad2008modelling}.
\citet{Helling_Mokhtarian_2001} argues that this is because flexible work makes accessibility less important in choosing a housing location. 
Decentralization has not been borne out thus far; urbanization has generally continued in developed and developing countries even as the share of flexible work has risen in recent decades. 
It could be that other factors have emerged that exert a stronger pull towards urban life, or perhaps flexible work as a share of the economy has yet to reach the activation energy required for the decentralization effect to take hold. 

Flexible work may reduce the demand for traditional offices and the demand for travel during peak hours, freeing up commercial space and transportation infrastructure for alternative uses \citep{Yu_Burke_Raad_2019}.
A survey of workers in Montreal, Canada found that those working in the central business district pre-COVID expect the highest degree of flexible work in the future, suggesting that demand for prime urban office space could be reduced \citep{shearmur2021towards}.
\citet{rosenthal2021jue} finds similar effects by studying U.S. commercial real estate prices, noting a decline in the premium for centrally-located office space.
The authors also point out that the effects are larger in dense, transit-oriented cities than in car-oriented cities.
On the infrastructure side, experts have argued for prioritizing safe and sustainable modes such as cycling and public transport rather than returning to the pre-pandemic status quo of private vehicle dominance \citep{Beck_Hensher_2021}.

A shift in working arrangements towards co-working spaces could provide new benefits, especially in smaller communities.  
\citet{mariotti2021geography} surveys co-working space users in Italy and finds that 85\% believe the co-working space has a positive impact on their community by hosting cultural events, purchasing from local services and improving security. 
These impressions were stronger for co-working spaces located outside of major metropolitan areas.
Policymakers in certain European countries had begun to promote co-working spaces in peripheral areas in order to boost economic development and reduce commutes \citep{mariotti2021research}.

\subsection{Organizational behavior} \label{sec:org_beh}

Organizational behavior is a critical component of flexible work. 
As stated by \citet{brewer2002flexible}, ``Individual work arrangements are influenced by the internal constraints of organisations such as organisational structure and managerial policies, including human resource management, work organisation, and external constraints such as union policies and agreements.''
The authors go on to provide an excellent overview of the ways in which organizational behavior affect travel behavior, and offer a call to action for additional research into the links between the two fields.
Their work builds on previous work by \citet{Brewer_1998}, who notes that flexible work requires both \emph{willingness} and \emph{capacity} from employers.  
Technological advances since the publication of that research may have increased the capacity of employers to facilitate flexible work, but willingness remains mixed.
A 2001 study of employers in Belgium found that the barriers to flexible work adoption were largely institutional rather than technological \citep{Illegems_Verbeke_SJegers_2001}. 

\citet{Handy_Mokhtarian_1996a} provides a good summary of early corporate resistance to flexible work, which includes concerns about productivity, supervision and morale loss due to ``officelessness''.
In an interesting reversal of the flexible work choice models described earlier, \citet{yen1994employer} use a survey to elicit the factors that contribute to \emph{employers'} decision to approve flexible work.  
The study finds that employers desire a smaller amount of flexible work than their employees prefer, much like the post-pandemic surveys issued nearly 30 years later. 
The primary concerns from employers were employee productivity, morale and communication. 

Empirical studies of flexible worker attitudes have shown a number of interesting relationships. \citet{Koh_Allen_Zafar_2013} find that flexible workers generally perceive higher support for work-life balance from their employer.
A meta-analysis of studies on flexible work \citep{martin2012telework} and organizational outcomes found that flexible work is perceived to ``increase productivity, secure retention, strengthen organizational commitment, and to improve performance within the organization.'' 
\citet{Coenen_Kok_2014} confirmed these results in a subsequent study.
\citet{Girit_2013} conducted a survey of flexible and traditional workers to compare personality traits, attitudes and performance. 
The survey found that flexible workers had higher performance scores and job satisfaction than those who worked in an office full time. 
Surprisingly, flexible workers considered themselves to be more extroverted than office-based workers, despite preferring a working arrangement with less face-to-face interaction. 
Studies conducted during the COVID-19 pandemic find that the relative productivity of flexible workers is dependent on job characteristics and the suitability of the home environment for work \citep{galanti2021work, george2021supporting}.

On the employee choice side, \citet{Mokhtarian_Bagley_2000} find that personal benefits and work effectiveness were two significant motivating factors in the decision to choose a workplace between home, a remote work center and a primary office.
Similarly, \citet{Laumer_Maier_2021} shows that household characteristics, including the suitability of the home for work activities, is an important determinant in the decision to work from home or elsewhere. 
\citet{Shafizadeh_Niemeier_Mokhtarian_Salomon_2007} examines the conditions under which flexible work is advantageous for the employee, the employer and society writ large. 
The authors find that flexible work is a net positive for employers when employee productivity does not diminish due to flexible work and when employers are able to translate flexible work policies into real estate savings. 
\citet{Bernardino_Ben-Akiva_1996} uses a comprehensive model of both employer and employee considerations to determine that flexible work has the potential to improve employee lifestyle and productivity, but less potential to reduce employer costs. 
\citet{choudhury2021work} finds that workers with the flexibility to relocate (``work from anywhere'') are more productive than flexible workers who visit a central office semi-regularly.  

In an extremely comprehensive review of the literature related to ``virtual work'', \citet{Raghuram_Hill_Gibbs_Maruping_2018} identifies a tremendous number of research gaps related to flexible work arrangements, teamwork and technology. 
The authors create a citation map of existing literature in these areas, finding that ``telecommuting'' research is a distinct cluster containing very few co-citations with other organizational behavior articles.
Studies related to the effects of flexible work on employees and their employers are summarized; employees with flexible work arrangements generally feel more empowered and less stressed, but also more isolated from their colleagues and less self-identification with their organization.
A case study of Microsoft employees working remotely before and after the onset of the COVID-19 pandemic found that collaboration became more static and sharing information became more difficult after the switch to remote work \citep{yang2021effects}. 

Co-working spaces have recently become a flexible option for employers who are growing rapidly or who prefer not to sign a long term lease.
As opposed to working at home, co-working spaces do not require investment in a home office, avoid the potential distractions of home-based work and allow interaction with colleagues or people from other organizations. 
\citet{Ross_Ressia_2015} and \citet{gandini2015rise} review the literature on co-working from an organizational behavior perspective, including the positive externalities of idea flow between co-located firms. 

One of the key limitations of the state of the art is understanding how and why people choose between alternative locations for flexible work (i.e. at home, a co-working space, a caf\'e). 
Furthermore, preferences for flexible work associates are not well studied. 
It has been shown that social relationships developed at work can affect commuting patterns \citep{plyushteva2019commutes}, so it is likely that these relationships would also affect location choices if multiple alternatives are available.

\section{Taxonomy} \label{sec:taxonomy}

Evidently there is a large body of research into flexible work that predates the COVID-19 pandemic. 
\citet{Raghuram_Hill_Gibbs_Maruping_2018} points out, however, that ``Telecommuting, by definition, includes satellite offices, telecenters, and client offices [...], but research in this area predominantly focuses on two locations: in-office versus not-in-office.''
Similarly, \citet{Pouri_Bhat_2003} notes that ``While home-based telecommuting is the dominant form of telecommuting today, insights into other forms of telecommuting, such as regional center telecommuting and neighborhood center telecommuting, are essential to the reliable estimation of overall telecommuting impacts.''
One could argue that even the expanded location examples given above are merely a small subset of possible workplaces. 
Furthermore, they ignore an entire dimension of work that has a significant impact on utility, productivity and travel: the people with whom one chooses to co-locate (or ``associates''). 
In this section, a formal taxonomy is proposed to capture the wide variety of possible working arrangements by combining different locations and associates. 
Then, the relationships between locations, associates, and other dimensions of work are enumerated and classified.
Such relationships will be referred to as ``dependencies''. 
Finally, the location-associates-dependencies taxonomy is used to propose and enumerate possible collaborations and business models that may arise to serve the mobility and workplace needs of flexible workers and their employers.
This is intended to provide future researchers with a common vocabulary for describing different work arrangements and their relationships between them for the purpose of transportation modeling.
It differs from previous flexible work taxonomy and framework development efforts \citep{mokhtarian1991defining,yeraguntla2005classification, Asgari_Jin_2015} in that it focuses on the locations and stakeholders as well as the possible relationships between them, with an explicit emphasis on the analytical modeling applications that are becoming critical tasks given the recent phase shift observed in flexible working trends. 

\subsection{Flexible work arrangements} \label{sec:work_arrangements}

Conceptually, each flexible work arrangement is a combination of one location and one or more associate groups, each of which have an impact on personal utility (i.e. satisfaction), productivity and transportation demand.
There are virtually infinite combinations; only the most common options for each category will be listed here.
Note also that while locations are mutually exclusive, associates are not. 
A colleague may also be considered a friend, for example. 

Locations will be described first.
As mentioned earlier, most flexible work research to date focuses on the two most common workplaces: the employer's designated workplace and home. 
There are many other options, however; some designed specifically for work and others that have a different primary purpose but facilitate work. 
The terms ``exclusive'' will be used to refer to an employer's workplace that does not involve any other organizations working in the same space, while ``shared'' indicates that the workplace is shared among several organizations. 
The following is a non-exhaustive list containing locations where flexible workers may choose to spend their working hours:

\begin{multicols}{2}
\begin{itemize}
    \item Primary employer workplace (exclusive)
    \item Primary employer workplace (shared)
    \item Satellite employer workplace (exclusive)
    \item Satellite employer workplace (shared)
    \item Client workplace
    \item Home
    \item Associate's home
    \item Co-working space 
    \item Caf\'e
    \item Library
    \item Community center
    \item Public outdoor area
    \item Hotel or vacation home
\end{itemize}
\end{multicols}

The distinction between the primary and satellite employer workplaces is that the staff at primary workplaces are grouped based on organizational structure, whereas the staff at a satellite workplace are grouped based on geography.
Not all of these options would be available to every flexible worker; for example, their employer may only have one office in the local region.
During the COVID-19 pandemic, many public and collaborative spaces were unavailable due to restrictions on businesses. 
Now that those restrictions are beginning to ease, many more working arrangements are possible.

In addition to location, work arrangements are differentiated based on co-located associates. 
While many people would choose work associates based on some relationship to their work-related needs, it is also conceivable to choose associates in order to socialize or for some non-work benefit.
The following list contains several categories of associates based on their relationship to the individual worker:

\begin{multicols}{2}
\begin{itemize}
    \item Colleagues
    \item Clients
    \item Family
    \item Friends
    \item People with the same profession
    \item People who work in the same industry
    \item No associates
\end{itemize}
\end{multicols}

These locations and associates can then be combined in different ways to produce work arrangements.
The dominant working arrangement among professionals during the second half of the 20th century was \emph{Colleagues + Primary employer workplace (exclusive)}. 
During the COVID-19 pandemic, many became familiar with either the \emph{No associates + Home} or \emph{Family + Home} arrangement.
As the nature of work and the economy continue to evolve, new arrangements might emerge.
One could imagine coordinating with friends to work out of each others' homes on a semi-regular basis (\emph{Friends + Associate's home}). 
Professional organizations might host a monthly event for flexible workers (\emph{Co-working space + People with the same profession}) to promote networking and skill transfer. 
So-called ``digital nomads'' may spend the majority of their time working while traveling, i.e. \emph{No associates + hotel}. 
Not all combinations of associates and locations are likely to occur, however.
For example, \emph{Friends + primary employer workplace (exclusive)} would be unusual, at least for friends who are not colleagues. 
A map of feasible work arrangements, visualized as a bipartite graph, is presented in Figure~\ref{fig:arrangements}. 
For simplicity, possible locations are categorized into five groups: home, associate's home, employer workplaces, co-working spaces and third places. 

\tikzstyle{block} = [draw, fill=white, rectangle, 
    minimum height=2em, minimum width=10em, text width=10em, text centered]

\begin{figure}[!ht]
\centering
\begin{tikzpicture}[auto,scale=0.8,node distance=1.2cm,>=latex']
    \node[block, name=home]{Home};
    \node[block, name=associate, below of=home]{Associate's Home};
    \node[block, name=workplace, below of=associate]{Employer Workplace};
    \node[block, name=coworking, below of=workplace]{Co-working Space};
    \node[block, name=third, below of=coworking]{Third Place};
    \node[block, name=friends, left of=home, xshift=-15em]{Friends \& Family};
    \node[block, name=colleagues, left of=associate, xshift=-15em]{Colleagues \& Clients};
    \node[block, name=prof, left of=workplace, xshift=-15em]{Same Profession};
    \node[block, name=industry, left of=coworking, xshift=-15em]{Same Industry};
    \node[block, name=none, left of=third, xshift=-15em]{None};
    \draw [-] (none.north east) -- (home.south west);
    \draw [-] (none) -- (third);
    \draw [-] ([yshift=-7.5pt]none.north east) -- (coworking.south west);
    \draw [-] (industry.south east) -- ([yshift=-11.25pt]third.north west);
    \draw [-] (industry) -- (coworking);
    \draw [-] (industry.north east) -- ([yshift=3.75pt]home.south west);
    \draw [-] ([yshift=-7.5pt]industry.north east) -- (associate.south west);
    \draw [-] (prof.south east) -- ([yshift=-7.5pt]third.north west);
    \draw [-] ([yshift=10pt]prof.south east) -- ([yshift=-10pt]coworking.north west);
    \draw [-] (prof.north east) -- ([yshift=7.5pt]home.south west);
    \draw [-] ([yshift=-10pt]prof.north east) -- ([yshift=7.5pt]associate.south west);
    \draw [-] (colleagues.south east) -- ([yshift=-3.75pt]third.north west);
    \draw [-] ([yshift=5pt]colleagues.south east) -- ([yshift=-5pt]coworking.north west);
    \draw [-] (colleagues.north east) -- ([yshift=11.25pt]home.south west);
    \draw [-] (colleagues) -- (associate);
    \draw [-] ([yshift=10pt]colleagues.south east) -- ([yshift=-15pt]workplace.north west);
    \draw [-] (friends.south east) -- (third.north west);
    \draw [-] ([yshift=7.5pt]friends.south east) -- (coworking.north west);
    \draw [-] (friends) -- (home);
\end{tikzpicture}
\caption{Illustration of the location - associate pairs that be combined to create feasible flexible work arrangements.}
\label{fig:arrangements}
\end{figure}
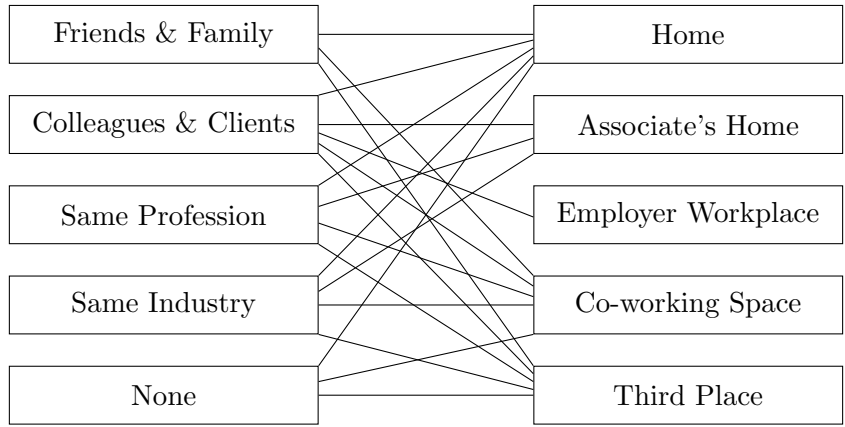

In reality, flexible workers may not choose one arrangement, but a set of possible arrangements and relative frequencies for each.
While some may prefer permanent flexible work, hybrid work schedules involving some flexible work and some face-to-face interaction appear to be the most popular arrangement among recent survey respondents, as noted in Section~\ref{sec:trends}. 
An example of a typical work week could be three days at a primary employer workplace and two days at home alone.
There may be week-to-week variation as well. 
Imagine someone with a hybrid schedule also works at a friend's house for one day every two weeks, and on occasion decides to work from a library rather than home for a change of scenery. 
These arrangements could even change throughout the course of a single day, with a flexible worker traveling to different locations or being joined by different associates. 

Each of these arrangements will result in a certain amount of utility for the flexible worker. 
Location and associates could also influence productivity, although the effect is nuanced and not necessarily a linear combination of influences.
For example, the benefit of working with colleagues would only accrue in full if the selected location supports collaboration. 
Furthermore, each of these arrangements affects the demand for travel in different ways.
Working with friends at a library could create additional pedestrian trips within a residential neighborhood, while working with people in the same industry at a downtown co-working space could create a mix of transit and private vehicle demand.

\subsection{Dependencies}

The relationships between a flexible worker, their location and their associates are not typically considered in existing transportation models. 
For convenience, these relationships will be referred to as ``dependencies''.
Dependencies generally originate on the demand side of the transportation modeling framework.
Imagine that a group of colleagues decide to work together on a given day.
That decision establishes a dependency: each of the colleagues must work in the same location. 
Then, the group chooses a location and individual travel arrangements in order to satisfy that dependency, based on some combination of preferences.
This is not a far-fetched scenario; a recent Harvard Business Review argued that ``allowing hyper-local teams to choose a location based on their shared preference'' is one of the many advantages of flexible work \citep{Choudhury_2020}. 

There are three categories of dependency: associate dependencies, geographic dependencies and facility dependencies.
A facility dependency is a requirement for a location with specific features or equipment, such as a worker who requires a workplace with a 3D printer in order to complete a design prototype. 
Geographic dependencies are related to the specific location of a workplace. 
An example of a geographic dependency could be someone who must choose a workplace in a certain neighborhood in order to pick up their children after school.
Associate dependencies are related to the need for co-location with specific individuals, perhaps two colleagues who desire face-to-face collaboration in order to finish a brainstorming task. 
These dependencies can also be combined.
If the two colleagues with an associate dependency also need a workplace with a conference room and a whiteboard, there is also a facility dependency.
Both dependencies affect their location choice set and decision process.

Dependencies can also be classified as either ``hard'' or ``soft''.
The 3D printer facility dependency above is a hard dependency because it is a strict requirement.
On the other hand, soft dependencies are desirable but are not strictly required.
Perhaps working in the same location as several friends provides social benefits but their presence is not required for the purpose of work.
Dependencies can also be chosen by the flexible worker (``bottom-up'') or enforced by the employer (``top-down''). 
Finally, dependencies can also vary by time of day, day of the week, or by task.

Facility and geographic dependencies can be incorporated relatively easily into destination choice models.
If they are hard dependencies, the set of available alternatives is limited to those that meet the conditions of the dependency.
If they are soft dependencies, then the presence of the facility type or the geographic location can be included in the utility function for each alternative workplace. 
Modeling associate dependencies, however, is more complex, as they introduce correlation between the decisions of multiple individuals. 
Improving discrete choice models to handle group decision making is one of the many research needs discussed in Section~\ref{sec:research}.

While these dependencies originate from the demand side of the transportation market, there are many transportation modes that are demand-responsive and would therefore benefit from internalizing flexible work dependencies. 
This is true for other flexible work stakeholders as well, such as co-working providers and even employers.
For example, an on-demand mobility platform could provide better workplace recommendations and travel options to a flexible worker if they could incorporate the worker's various dependencies. 
Similarly, a co-working provider could offer customized subscription bundles to flexible workers based on their facility preferences, geographic constraints and desire to work with others. 
Incorporating these dependencies into supply optimization models can be challenging, and is discussed further in Section~\ref{sec:research}.
The interactions between these stakeholders and opportunities for collaborative business models are described in the next section. 

In summary, there are three dependency categories: associate, geographic and facility.
Furthermore, dependencies can be either hard constraints or simply desirable, and enforced from the top-down or bottom-up. 
From a practical perspective, dependencies can be incorporated into demand models in different ways, and into demand-responsive supply models, depending on their properties.

\subsection{Stakeholders} \label{sec:stakeholders}

The relatively small incidence of flexible work leading up to 2020 meant that there was a limited market for services and platforms to support variable working arrangements.
Since then, however, there have already been a number of new offerings, business models, and partnerships.
Many more are likely to arise in the future as the market for flexible work services continues to grow.
As described in previous sections, flexible work involves many decisions on the part of workers and employers, creating an opportunity for services that provide integrated workplace and mobility solutions.
Workers and employers may have different preferences, however. 
Additionally, governments will play an important role as a regulator and service provider with an interest in public welfare.
In this section, each of the primary stakeholders for flexible work are described, identifying decision makers and the incentive structures that promote cooperation. 

A typical flexible work arrangement involves five key stakeholders which may or may not be separate entities. 
First, of course, is the flexible worker themselves, who chooses an arrangement from the available alternatives and a complex set of preferences.
Second is the employer, who defines the set of available alternatives through their flexible work policies, and who may seek to influence the worker's choice of arrangement with additional compensation or direct reimbursement.
Employers could have many motivations related to flexible work, but among the most common are employee performance, employee morale, and expenses.
The workplace operator is the third stakeholder, whether that is a traditional commercial real estate lessor, a co-working platform, a caf\'e or even an employer-owned property.
These workplace operators are primarily motivated by financial incentives.
Fourth is the mobility provider.
If the flexible work location is outside the home, then different providers may compete for the trip on the basis of cost and service quality.
Even the traditional suburb-downtown commute can have a variety of travel options: drive alone, carpool, car-sharing, park-and-ride, employer shuttle, taxi, commuter rail, and so on.
Finally, policy makers can play a role by encouraging certain arrangements through incentives or regulation. 

Note that the five key stakeholders enumerate above do not constitute an exhaustive list, there are many other stakeholders affected by flexible work arrangement choices.
This includes small businesses who cater to office workers, captive transit riders who are affected by changes in service, and so on.
These constituencies are important to consider in policy decisions, but are not directly involved in determining the feasibility of different workplace arrangements. 

The following extended example will illustrate the possible interactions between stakeholders.
Imagine a flexible worker whose employer allows them to choose their own working arrangement multiple days per week.
The worker decides to spend every Thursday at a co-working space a few miles from their home alongside several people who work in the same profession.
This choice of working arrangement is entirely the prerogative of the worker (Stakeholder 1), given their employer's flexible work policy (Stakeholder 2).
Next, the worker will need to choose a specific co-working location (Stakeholder 3) in coordination with their intended associates.
The worker may choose a location based entirely on their own preferences and transportation concerns, or their employer could subsidize (but not mandate) a certain set of co-working spaces because the employer believes that their employees are more productive and content when working around others.
This is one example of the second and third stakeholders working together for mutual benefit; not surprisingly, such partnerships have become commonplace over the past decade \citep{bacevice2019}.
Alternatively, the employer might provide a direct payment for co-location of colleagues. 
The market research company Ask Your Target Market covers long-distance travel expenses to encourage face-to-face meetings between their flexible workers, as well as the cost of lunch for employees who spend the day working together \citep{aytm2021}.

Next, the worker will have to choose a travel mode (or modes) for their trips to and from the workplace, thus engaging with a fourth stakeholder.
They might choose to drive or bike, or to use public transit.
Perhaps the employer provides a free transit pass to encourage employees to choose an out-of-home working arrangement where there are fewer distractions, thus creating a relationship between the employer and mobility providers.
Alternatively, the employer may contract with a ride-hailing platform to provide subsidized trips for their employees to and from co-working spaces for the same purpose.
In the second case, the ride-hailing platform and co-working space could also partner to create an integrated office space and mobility service to offer to individuals and employers.
Fully integrated services have not been developed to date, but there are some examples of mobility and workplace providers working together; for example, WeWork offers a monthly Uber Pass as part of their ``All-Access'' subscription \citep{kunesh2021}.
These partnerships are not limited to on-demand mobility, either. 
Workplace providers could also offer discounted transit passes, bikesharing or carsharing subscriptions, or even operate their own shuttles to pick up members in the surrounding area.
Each partnership results in a business relationship between the mobility and office services with some revenue sharing agreement. 

Finally, public institutions play both a direct and indirect role by allowing certain land uses, operating infrastructure, and licensing private businesses. 
Local agencies could encourage sustainable mobility decisions through congestion pricing or providing public transit services that are oriented around flexible work.
Zoning regulations might be designed to promote co-working spaces and satellite offices in residential areas, affecting the available locations for flexible work.
Each of these policies could be crafted with direct input from workers, employers, mobility providers or workplace operators, thus creating a set of regulatory and advocacy relationships with each of the other flexible work stakeholders. 

\begin{figure*}[ht]
    \begin{center}
        \includegraphics[width=0.7\textwidth]{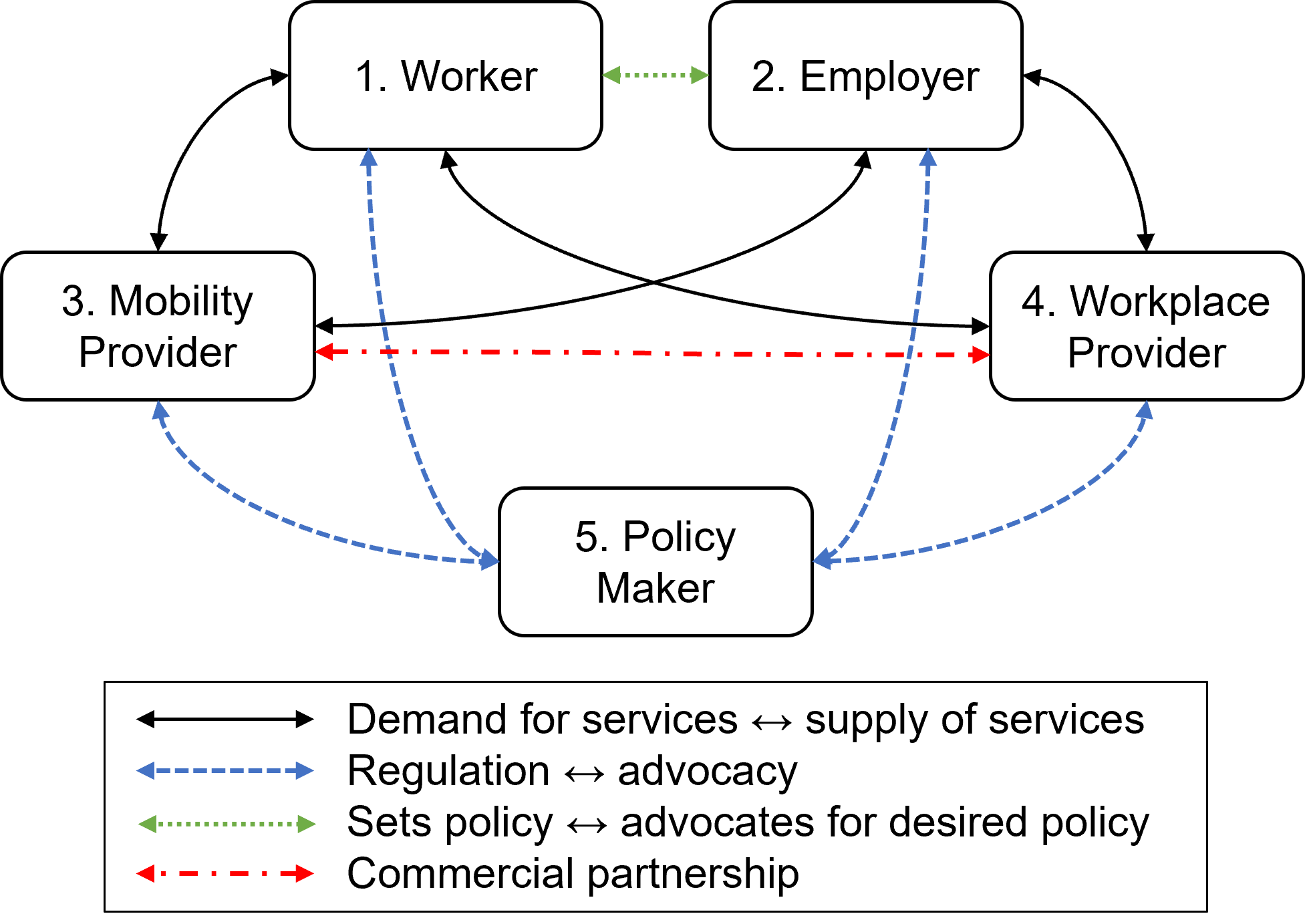}
        \caption{Flexible work stakeholders and relationships.}
        \label{fig:stakeholders}    
    \end{center}
\end{figure*}

The five key flexible work stakeholders and the possible relationships between them are summarized in Figure~\ref{fig:stakeholders}.
Certain pairs of stakeholders have a supply - demand relationship, such as workers and mobility providers. 
Workers seek mobility services and providers offer those services. 
Policy makers interact with other stakeholders by issuing regulations and incentives, while the stakeholders advocate for their preferred policies. 
Two unique relationships exist: 1) employers set flexible work policies for workers, with or without input from the workers themselves, and 2) mobility providers and workplace providers may partner together to offer service bundles. 

Continuing the example above, imagine that an employer has a significant number of flexible workers in a given area, and rather than coordinating their work locations, the employer contracts with a hybrid workplace-mobility company that provides each worker with access to a network of co-working spaces and a bundle of mobility services.
The employer pays this hybrid company a flat monthly subscription fee per employee. 
Each work day, the flexible worker uses a mobile application to choose from a set of alternative workplace and travel options. 
To return to dependencies briefly, this service might internalize the needs of the employer's flexible workers by allowing them to specify certain types of facility or geographic dependencies before choosing an alternative.
Moreover, the service might also allow the user to specify one or more associates for co-location.
The desire to co-locate with team members may be a personal preference, or perhaps the employer offers each worker a small financial incentive to encourage face-to-face interaction due to the perceived productivity benefits. 
Given all of these connections between stakeholders and dependencies, the hybrid workplace-mobility provider must then solve a very complex assignment and routing task in order to serve each customer while ensuring efficient allocation of their own resources.
Public policies that limit negative transportation externalities, such as peak hour road pricing, add yet another layer of complexity to the assignment task.
This is just one example of a new analytical modeling problem, described in the language of flexible work arrangements and dependencies, that must be solved to enable sustainable mobility for the future of work. 
The next section presents a conceptual framework for such problems. 
\section{Model classification} \label{sec:framework}

Having established a list of possible flexible working arrangements, the dependencies that affect the demand for these arrangements, and the stakeholders who determine the outcomes, the next step is to classify the resulting modeling problems.
Broadly speaking, four of the stakeholders described in the previous section could benefit from designing and solving system-level optimization problems that affect multiple individuals: employers, workplace operators, mobility providers and the public sector. 
Given that this paper is primarily concerned with the mobility impacts of the future of work, the focus of this section is on analytical models for mobility providers, although considerations for the other stakeholders are included as research needs in Section~\ref{sec:research}.

Adapting transportation supply optimization problems for the future of work involves on understanding the day-to-day location choices of flexible workers, and using that information to improve service. 
Each class of transportation mode has many models for optimizing service delivery, and each of these models interacts differently with flexible dependencies.
This framework can be represented conceptually as a table with supply models on the vertical axis and future of work characteristics on the horizontal axis, as shown in Table~\ref{tab:framework}. 
Flexible locations are included as the first future of work characteristic because many existing transportation models assume that the work destination is fixed.
Travel modes, including autonomous vehicles (AVs), are divided into three broad categories based on models of ownership: public, shared, and private. 
Each of the cells represents one or more possible supply-demand models that captures the interaction between mobility and flexible work.

\begin{table}[ht!]
    \begin{tabular}{|m{0.06\textwidth} m{0.21\textwidth} | m{0.15\textwidth} | m{0.14\textwidth} | m{0.14\textwidth} | m{0.14\textwidth} | }
    \hline
         \multicolumn{2}{|l|}{\multirow{2}{*}{\textbf{Travel Mode}}} & \multicolumn{4}{c|}{\textbf{Future of Work Characteristics}} \\ \cline{3-6}
         & & \textbf{Flexible Destinations} & \textbf{Associate \newline Dependency} & \textbf{Geographic \newline Dependency} & \textbf{Facility \newline Dependency} \\ \hline
         \multirow{2}{*}{Public} & Fixed Route Transit &  &  &  & \\ \cline{3-6}
         & Flexible Transit &  &  &  & \\ \hline
        \multirow{4}{*}{Shared} & TNCs + Taxis &  &  & & \\ \cline{3-6}
         & Ride-pooling & & & & \\ \cline{3-6}
         & Shared AVs &  &  & &  \\ \cline{3-6}
         & Micromobility &  &  &  & \\ \hline
         \multirow{4}{*}{Private} & Active travel &  &  & & \\ \cline{3-6}
         & Private gas car &  &  & & \\ \cline{3-6}
         & Private electric car &  &  & & \\ \cline{3-6}
         & Private AV &  &  & & \\ \hline

    \end{tabular}
    \caption{Conceptual table for the mobility and future of work analytical framework}
    \label{tab:framework}
\end{table}

The supply models on the vertical axis are very diverse.
A public transit supply model with flexible work locations could involve setting bus schedules that vary across weekdays because people tend to work at home more often on Mondays and Fridays.
On the other hand, a ride-pooling model with flexible locations could involve recommending alternative destinations to customers in order to improve the likelihood of serving multiple trips with a single vehicle.
Positioning future transportation research within this tabular framework will help investigators understand the similarities and differences between their work and existing models.

It should be noted that Table~\ref{tab:framework} is not an exhaustive list of travel modes, but rather some general categories.
Furthermore, the cells are not intended to be mutually exclusive. 
Several examples of ongoing research projects are included in the next subsection to demonstrate how this framework can be applied.

\subsection{Examples}

This section applies the framework developed above to several examples of ongoing research projects exploring different modes and future of work characteristics.

The first example is the design of a transit-centric multi-modal mobility (TCMM) system that incorporates shared mobility as a first and last mile mode for accessing transit stations. 
Flexible work may result in more trips within or between residential areas and fewer commuting trips to the downtown core. 
Joint design of a shared mobility system and transit network to serve demand with flexible destinations could improve service levels, thus making sustainable travel a more attractive option for flexible workers. 
This example occupies multiple cells in Table~\ref{tab:framework}; it falls within the flexible destinations column, and covers each of the public and shared modes as well as active travel. 

The second example is the development of new ride-pooling algorithms that can internalize both destination flexibility and the dependencies that govern the choice set of travelers.
By adapting canonical matching algorithms to include location-specific and agent-specific variables, associate dependencies, geographic dependencies and facility dependencies can be strictly enforced or included as penalty terms. 
Unlike the previous example, which occupied a single column and multiple rows, this example occupies each of the columns along the ride-pooling row.

A third example includes the development of zoning and co-working site selection models.
The primary research question is how to distribute and size co-working spaces efficiently with respect to some objective, such as minimizing commuting distance or maximizing revenue.
This distribution is subject to constraints, which might include facility and geographic dependencies for a subset of the potential demand. 
If driving was assumed to be the primary commuting mode, then such a model would occupy a combination of rows and columns in Table~\ref{tab:framework}: the private gas and electric car modes along with geographic and facility dependencies. 

The examples of methodological research provided in this section, from improved ride-pooling algorithms to mixed-use zoning allocation, would each benefit from updated evidence and new theories regarding flexible work. 
Specific research gaps and needs are described in the next section.

\section{Research needs} \label{sec:research}

Adapting to the future of work is not simply a transportation problem, although transportation is intrinsic to all flexible work decisions.
Employers, workplace providers and governments are currently operating with considerable uncertainty about the future of work and would benefit from an optimization framework that can be used for data-driven decision making.
For example, employers could use incorporate employee preferences, home locations and productivity functions when designing flexible work policies and incentives.
Workplace operators can leverage trip record data for locating and sizing new facilities, and for setting subscription prices. 
Governments would benefit from a strategic approach to planning flexible work land uses in order to spur economic growth in struggling neighborhoods or to mitigate congestion within the transportation network. 
Each of these applications requires a robust source of data as well as new methodologies that can incorporate flexible work dynamics.
For that reason, flexible work research needs described in this section are categorized into two important subsets: empirical research and methodological research. 

\subsection{Empirical research}

While the long term trend towards flexible work has yet to be observed due to ongoing concern about the COVID-19 pandemic and related economic restrictions, it is not too early to begin to collect empirical evidence.
Empirical research needs can be further categorized into three major themes: 1) Changes to individual behavior, 2) Employer policies, and 3) Flexible work infrastructure. 
These empirical research goals are summarized by theme at the end of this subsection.

Surveys of individual behavior have been conducted throughout the course of the pandemic, as outlined in Section~\ref{sec:trends}, including many that ask about future intentions.
Unfortunately the methodologies of these surveys are inconsistent, making it difficult to compare them against one another.
The first primary research goal should be to develop a standardized set of survey questions that would allow comparison over time, between different geographic areas and between socioeconomic groups.

Second, in order to predict long term trends, surveys of both employers and workers should be continued on a regular basis.
Typical sources of socioeconomic and travel behavior data, such as the American Community Survey and the National Household Travel Survey, are conducted infrequently and thus not suitable for understanding the rapidly changing dynamics of flexible work.
Attitudes towards the pandemic are in flux due to the emergence of virus variants on one hand and continued vaccination progress on the other, therefore attitudes towards flexible work might also evolve.
Panel studies or updated versions of previous studies would allow researchers to study how these attitudes have changed over time and to associate these changes with the dynamics of the pandemic.
The results could also provide insight into the temporal evolution of general behavioral trends during public health crises. 

Next, empirical studies should expand their scope beyond the immediate questions around flexible work to include detailed questions about preferred working arrangements, housing choices, commuting and non-work travel.
With the exception of working arrangements, these themes have been included on a handful COVID-19 era surveys, but rarely in sufficient detail to support modeling efforts or policy development.
Eliciting the factors that influence the decision by flexible workers to move home locations, to start cycling to work, or to switch to a different grocery store, will have important implications for transportation services, real estate markets and public policy.
The extent to which employer-provided perks such as free on-site parking or transit passes have changed after the pandemic, and how those perks affect flexible work decisions, is another potential research topic. 

To the authors' knowledge, no surveys have asked the large cohort of workers who expect to engage in flexible work after the pandemic about their preferences for different work arrangements. 
Determining whether this group prefers flexible work to take place at home alone, at a caf\'e with friends or a co-working space with colleagues is one of the most critical research needs in this area.
Well-designed stated choice surveys of flexible workers would allow researchers to infer the willingness-to-pay and willingness-to-travel measures for different workplaces and associates.
Comprehensive studies of these preferences, contributing factors, and population heterogeneity could be used to inform transportation planning, employer policies and investment decisions, among many other decisions. 

Stated choice surveys are helpful, especially when forecasting future trends, but there are also opportunities to collect and analyze revealed preference data related to flexible work. 
Established methods for tracing mobility patterns, such as mobile phone traces and transit smartcards, can and have been used to study changes in demand for travel during the COVID-19 pandemic \citep{franch2020spatial}.
Datasets that allow for longitudinal comparison of individuals before and after COVID-19 restrictions are especially valuable, as they permit quantification of the spatial and temporal changes in work activities. 
New methods could be applied to revealed choice data, such as credit card transactions and mobility patterns, to estimate demand functions for different workplace types. 
Machine learning and statistical inference techniques might leverage these datasets to identify new patterns of behavior related to flexible work. 
This information would help transit agencies and other transportation providers adapt to changing demand patterns. 

One application of this preference data is in cost-benefit analysis for transportation projects.
A greater prevalence of flexible workers might change the determination for new transit services, bike sharing programs or road projects.
Previous calculations would have considered a low or moderate amount of induced demand for new transit projects as a result of travelers switching from other modes or changes in destination for some discretionary trips.
The destinations of commuting trips, however, are largely considered to be fixed \citep{litman2015evaluating}. 
Collecting empirical data about workplace preferences allows planners to model the changes in work locations that result from improvements to the transportation network, and would ultimately affect which projects are proposed and implemented.

Turning to the employer policy theme, the degree to which employers share the preferences of their employees, or have countervailing preferences is perhaps more important, as employers ultimately determine flexible work policies.
Much like mobility patterns, recent research around employee and firm productivity has focused on the period at the beginning of the COVID-19 pandemic when most flexible work was conducted at home \citep{galanti2021work, george2021supporting}.
Given that a much larger share of the population is expected to engage in flexible work in the future, it is important to understand how different work arrangements affect individual and organizational outcomes.
Rigorous case studies of employers who adopted widespread flexible work or distributed satellite offices before the pandemic, their organizational structures, flexible work policies and communication practices could be informative. 
Furthermore, the actions and stated intentions of major employers during the COVID-19 period related to work site relocation, flexible work policies and organizational restructuring should be cataloged. 

Finally, there is the issue of creating and curating new sets of data that can facilitate future of work travel demand models.
One important research need is the location, capacity and occupancy of potential workplaces for flexible work within different urban areas.
This would include each of the public and rentable facilities described in Section~\ref{sec:work_arrangements}, such as caf\'es, libraries and co-working spaces.
Previous studies of flexible work destinations, such as \citet{Ge_Polhill_Craig_2018}, have made strong assumptions about available workplaces due to a lack of ground truth data.
Going forward, an accurate representation of availability will be necessary to generate actionable insights.  

Table~\ref{tab:empirical} summarizes the empirical research needs described above:

\renewcommand{\arraystretch}{1.5}
\begin{table}[ht!]
    \begin{tabular}{|m{0.19\textwidth}  | m{0.75\textwidth} | }
    \hline
         \textbf{Theme} & \textbf{Research topic} \\ \hline
         \multirow{4}{*}[-1em]{Individual behavior} & 
         Establish a set of standard questions for future flexible work surveys and continue existing surveys throughout the course of the COVID-19 pandemic  \\ 
         & Expand surveys to include comprehensive questions about changes in lifestyle brought about by an increase in flexible work\\ 
         & Collect or synthesize \emph{revealed preference} data to study actual changes in mobility patterns \\ 
         & Focus surveys on preferences for different flexible work arrangements \\ \cline{1-2}
         \multirow{3}{*}[-1em]{Employer policies} & Survey employers across industries and compare difference in flexible work preferences between employers and staff \\
         & Track aggregate trends in workplace relocation, flexible work policies and organizational restructuring \\
         & Review the practices and outcomes of organizations who adopted extensive flexible work before March 2020 \\\cline{1-2}
        {Infrastructure} & Create new datasets identifying the location and capacity of available flexible workplaces \\ \hline

    \end{tabular}
    \caption{Summary of empirical research needs}
    \label{tab:empirical}
\end{table}

\subsection{Methodological research}

The collection of new empirical data will have limited impact if the dynamics of flexible work cannot be captured by analytical models. 
These models should address the need for long-term strategic planning by mobility providers, employers and workplace providers, but also operational decisions such as real-time dispatching and matching functions.
Many of these research needs are related directly to transportation, and thus fit within the framework described in Table~\ref{tab:framework}, although some additional opportunities for future organizational behavior and urban planning research are also proposed.
Similar to the empirical research, these needs have been organized into three categories: 1) Travel demand modeling, 2) Supply adaptation and 3) Land use and public policy. 
The travel demand modeling tasks will benefit from the individual and employer-focused empirical research described in the previous section. 
The supply adaptation research are then built on top of the demand modeling improvements, offering a response that caters to the needs of flexible workers.
Lastly, the land use and public policy needs incorporate all three themes of empirical research, including flexible work infrastructure. 

In many cases, these research needs involve adding new components or complexity to existing methods to reflect the opportunities and restrictions introduced by flexible work.
Other proposed areas of research require integration of multiple existing methods or models from different disciplines.
Finally, two of the research needs call for entirely new methods to quantify important flexible work dynamics that are known to exist but difficult to analyze effectively. 

First, associate dependencies imply a joint decision between multiple individuals, a scenario that has been historically difficult to encode into the discrete choice modeling framework. 
The flexible work location choice of one colleague could affect the decision of another, and so on; these effects should be captured within the choice model.
There have been notable efforts to include a single decision made by a multi-person household \citep{zhang2009modeling}, or decisions where social effects are incorporated \citep{brock2001discrete}, but models of day-to-day decision making with correlated outcomes are limited (for an example, see \citet{zemo2018farmers}).
These efforts could involve improving the traditional econometric models, or perhaps creating new choice models altogether.
For example, when a group of colleagues with flexible work are given the freedom to choose a time and place for a collaborative meeting, the negotiation between alternative destinations could be modeled as a set of strategic choices between decision makers, thus introducing game theory concepts into destination choice models. 
Machine learning for travel behavior modeling has been growing in popularity; such techniques may prove more suitable for the collective choice between flexible work arrangements.
Moreover, work arrangements are likely to be influenced by previous choices. 
Choice models should incorporate memory of previous decisions, and the competing desires between exploration of new possibilities and exploiting known work arrangements.

Modeling the choice of working arrangements necessarily includes a utility function for different destinations; empirical research described in the previous section can help to calibrate the function parameters. 
This is one example of adding new complexity to an existing method.
Many classical travel demand models \citep{ali1988combined, oppenheim1993equilibrium} do not include occupancy in the destination utility function, or require that destination utility is a monotonic function of occupancy in order to guarantee that a solution exists and that it is unique. 
Destination choices, including workplace choices, are often non-monotonic, however.
For example, the utility gained from working at a co-working space might increase with occupancy, but only up to a certain threshold, after which utility begins to decrease rapidly due to crowding and noise. 
Other, more complex functions are certainly possible.
Generalized travel demand models that accommodate such functions are therefore necessary for accurate estimates of transportation outcomes in the future of work.

Advances in the modeling of flexible work arrangements and destination utility could be leveraged for a new application: recommendation engines for flexible workers. 
Some of the stakeholder partnerships described in Section~\ref{sec:stakeholders} involve collaboration between mobility platforms and workplace providers to sell integrated services to workers or their employers. 
Methods for clustering different users based on their transportation and work arrangement preferences, and for learning how their preferences change over time, would help these partnerships recommend different service bundles to prospective and returning users. 

The research needs described above are largely related to demand modeling, but supply models must also be adapted for flexible work. 
One of the biggest concerns around flexible work is the overall uncertainty.
As described in Section~\ref{sec:trends}, there is much speculation around flexible work trends, and the majority of studies simply estimate flexible work in the near future, rather than five or ten years hence. 
Across all transportation modes, one critical research need is supply optimization models that allow for uncertainty and future changes in the quantity, location and timing of travel demand.
There are several methodological approaches that incorporate uncertain parameters, including stochastic optimization and robust optimization, which could be applied to transportation planning problems.
For example, transit network design models should be capable of including the possibility that flexible work grows to 40\% of worked hours in the long run as well as the possibility that it continues to account less than 10\% of worked hours.
These new techniques should then find solutions that work well in each of the possible scenarios, or solutions that are inherently flexible, resilient and adaptable to changing trends.

There has been considerable work over the past decade in developing efficient trip matching algorithms for ride-sharing, which includes ride-pooling operations, car-pooling schemes and demand-responsive transit.
These algorithms generally assume a fixed origin and destination for each trip, however \citep{alonso2017demand}.
If the customer is a flexible worker, they might consider several alternative work locations and seek to compare the travel costs for each.
In a similar vein, perhaps the customer is interested face-to-face meeting with several colleagues and is seeking a set of convenient flexible work locations. 
Creating new matching algorithms that capture these associate dependencies, as well as facility dependencies like capacity constraints, is an important research topic for shared mobility providers.

Much like ride-sharing, transit assignment modeling has typically assumed fixed destinations for most transit trips (e.g., \citet{oliker2018frequency}).
Flexible workers, on the other hand, may adjust their destinations each day based on the availability and quality of transit service, or even due to new information received en-route, such as the presence of downstream incidents. 
New transit assignment models are needed that can capture these dynamics, including the effect of real-time information, with tractable formulations to support transit planning applications.

Moreover, transit networks and schedules are generally designed around the stable spatio-temporal demand produced by fixed commuting patterns. 
New trends may emerge as people become more comfortable with flexible work; one might imagine that more flexible work will occur on Mondays and Fridays than mid-week, but that people are more likely to work in non-home locations on Fridays than Mondays in order to pursue social activities after work.
These trends have not been a major issue for transit agencies due to limited flexible work before March 2020, but could create significant disparities in aggregate travel demand between weekdays.
Adjusting to these patterns could require the creation of different transit schedules or even route patterns for each weekday, which is not common practice today.
Creating these individual timetables will require new methods for rapid data collection and schedule optimization, as well as research into dissemination and communication of complex timetables to the public and to transit operators.
Schedule optimization models could also be updated to endogenize the flexible work policies of major regional employers who have a substantial effect on the demand for public transit. 

Finally, there are clearly flexible work research needs in fields related to transportation planning such as organizational behavior and land use planning.
Practitioners in both fields, like those in transportation, are responding to rapidly changing conditions with a general lack of evidence and information.
Employers are currently tasked with setting flexible work policies for the post-pandemic period despite considerable uncertainty around how different policies will affect performance, morale and retention.
Methodological innovations that can provide an accurate link between these policies and their organizational outcomes are sorely needed; some may even include transportation components.

Productivity modeling presents the two opportunities for the development of entirely new methods to capture flexible work dynamics.
First, detailed methods for identifying and predicting the effects of different flexible work arrangements on attitudes such as trust between colleagues and self-identification with a career or employer, and relate these attitudes to organizational outcomes (e.g. productivity, retention). 
Second, ``knowledge spillover'' is an economic benefit that arises from interactions and discussions between people, even if those people are in different industries or professions \citep{glaeser2010agglomeration}. 
These benefits have been observed at a macro scale, but it is difficult to measure the conditions or policies that affect knowledge spillover when the interactions are often spontaneous and their effects can take some time to materialize.
As employers evaluate different flexible work policies, such as leasing space in a co-working environment or incentivizing employees to choose collaborative flexible work arrangements, it is critical to model the productivity benefits of these policies that result from interacting with others.
Adapting methods from network science to this problem is one promising direction of research. 

Land use planners are in a similar position to large employers.
Knowledge spillover can benefit the local economy, so designing policy and land use to encourage interactions between individuals should be prioritized, but new methods will be necessary to do so. 
There is little consensus on how flexible work will change the demand for different land uses, including infrastructure, in the long term.
Improved urban economic models that allow for new flexible work-oriented land uses, travel patterns and the impact on agglomeration effects should be developed to guide future decision making. 

Table~\ref{tab:methodological} summarizes the methodological research recommendations:

\renewcommand{\arraystretch}{1.5}
\begin{table}[ht!]
    \begin{tabular}{|m{0.19\textwidth}  | m{0.75\textwidth} | }
    \hline
         \textbf{Theme} & \textbf{Research topic} \\ \hline
         \multirow{4}{*}[-1em]{Travel demand} & 
         Incorporate correlation and negotiation between the destination choices of individuals within discrete choice-based travel demand models \\ 
         & Expand the exploration-exploitation trade-off for destination choice models to include the choice of workplace\\ 
         & Develop tractable travel demand models that permit non-monotonic destination utility functions \\ 
         & Explore new methods for synthesizing mobility and location choices in order to support a recommendation service for flexible workers \\ \cline{1-2}
         \multirow{3}{*}[-1em]{Supply adaptation} & Build supply optimization models that can handle uncertainty in flexible work trends \\
         & Create optimal passenger-vehicle matching algorithms that can capture associate, geographic and facility dependencies \\
         & Propose new transit and multi-modal network design and scheduling problems that involve irregular and flexible commuting patterns \\\cline{1-2}
         \multirow{2}{*}[-1em]{Land use \& policy} & Build dependencies into regional planning models to predict how new zoning regulations or developments will affect location choices and commuting patterns \\
         & Develop new methods for modeling the impact of flexible work arrangements on knowledge spillover \\
         & Create models for individual and firm productivity in a flexible work environment with dependencies, enabling evidence-based employer policies \\ \hline

    \end{tabular}
    \caption{Summary of methodological research needs}
    \label{tab:methodological}
\end{table}

\section{Conclusions} \label{sec:conclusion}

Urban mobility is facing a remarkable change in the underlying demand for travel due to the abrupt shift towards flexible work.
This new reality presents both challenges and opportunities for transportation professionals and researchers.
Old assumptions about stable commuting patterns and large, centralized workplaces are no longer valid.
To ensure that the future of work is also a future with sustainable and equitable mobility, transportation systems must adapt to the changing demand brought about by flexible work.
Coordinating a research effort around this crucial adaptation, however, is difficult without a definitive terminology and conceptual structure.

This paper addresses that issue by creating a taxonomy for describing work arrangements, their stakeholders, and the possible relationships between them.
A research framework is then proposed to organize, compare and contrast future efforts towards inserting flexible work dynamics into transportation models.
Both the taxonomy and the research framework are illustrated using a case study of a rapidly evolving neighborhood in Boston, Massachusetts.
Finally, seventeen empirical and methodological research needs are identified, with emphasis on topics that have immediate practical applications. 


The rise of new technology and global events have often precipitated large-scale shifts in travel behavior. 
We are living through a period where both are occurring simultaneously, presenting a limited window of opportunity to influence future transportation and land use outcomes.
In the 20th century, such opportunities ultimately resulted in automobile-dominant transportation systems and urban decentralization. 
This time, through effective policy leadership informed by thoughtful, coordinated research, we might finally take advantage of a liminal moment to shape a more sustainable future.

\section{Acknowledgements}

The authors would like to thank Jim Aloisi for his thoughtful feedback on an early draft of this paper. 

\bibliographystyle{apalike}
\bibliography{reference}

\begin{thebibliography}{}

\bibitem[Alexander et~al., 2021]{mckinsey2021}
Alexander, A., De~Smet, A., Langstaff, M., and Ravid, D. (2021).
\newblock What employees are saying about the future of remote work.
\newblock
  \url{https://www.mckinsey.com/business-functions/organization/our-insights/what-employees-are-saying-about-the-future-of-remote-work}.
\newblock Online; accessed on 2021-07-12.

\bibitem[Ali~Safwat and Magnanti, 1988]{ali1988combined}
Ali~Safwat, K.~N. and Magnanti, T.~L. (1988).
\newblock A combined trip generation, trip distribution, modal split, and trip
  assignment model.
\newblock {\em Transportation Science}, 22(1):14--30.

\bibitem[Alonso-Mora et~al., 2017]{alonso2017demand}
Alonso-Mora, J., Samaranayake, S., Wallar, A., Frazzoli, E., and Rus, D.
  (2017).
\newblock On-demand high-capacity ride-sharing via dynamic trip-vehicle
  assignment.
\newblock {\em Proceedings of the National Academy of Sciences},
  114(3):462--467.

\bibitem[Arabikhan, 2017]{arabikhan2017telecommuting}
Arabikhan, F. (2017).
\newblock {\em Telecommuting choice modelling using fuzzy rule based networks}.
\newblock PhD thesis, University of Portsmouth.

\bibitem[Asgari and Jin, 2015]{Asgari_Jin_2015}
Asgari, H. and Jin, X. (2015).
\newblock Toward a comprehensive telecommuting analysis framework.
\newblock {\em Transportation Research Record}, 2496:1–9.

\bibitem[Asgari and Jin, 2017]{Asgari_Jin_2017}
Asgari, H. and Jin, X. (2017).
\newblock Impacts of telecommuting on nonmandatory activity participation: Role
  of endogeneity.
\newblock {\em Transportation Research Record}, 2666:47–57.

\bibitem[Asgari et~al., 2014]{asgari2014choice}
Asgari, H., Jin, X., and Mohseni, A. (2014).
\newblock Choice, frequency, and engagement: Framework for telecommuting
  behavior analysis and modeling.
\newblock {\em Transportation Research Record}, 2413(1):101--109.

\bibitem[Asgari et~al., 2019]{Asgari_Jin_Rojas_2019}
Asgari, H., Jin, X., and Rojas, M.~B. (2019).
\newblock Time geography of daily activities: A closer look into telecommute
  impacts.
\newblock {\em Travel Behaviour and Society}, 16:99–107.

\bibitem[{Ask Your Target Market (aytm)}, 2021]{aytm2021}
{Ask Your Target Market (aytm)} (2021).
\newblock {Meet your teammates, both near \& far}.
\newblock \url{https://insights.aytm.com/aytm-careers}.
\newblock Online; accessed on 2021-08-17.

\bibitem[Bacevice et~al., 2019]{bacevice2019}
Bacevice, P., Spreitzer, G., Hendricks, H., and Davis, D. (2019).
\newblock How coworking spaces affect employees’ professional identities.
\newblock {\em Harvard Business Review}.

\bibitem[Bagley et~al., 1994]{Bagley_Mannering_Mokhtarian_1994}
Bagley, M.~N., Mannering, J.~S., and Mokhtarian, P.~L. (1994).
\newblock Telecommuting centers and related concepts: A review of practice.
\newblock Technical report, University of California, Davis.

\bibitem[Bagley and Mokhtarian, 1997]{bagley1997analyzing}
Bagley, M.~N. and Mokhtarian, P.~L. (1997).
\newblock Analyzing the preference for non-exclusive forms of telecommuting:
  Modeling and policy implications.
\newblock {\em Transportation}, 24(3):203--226.

\bibitem[Barrero et~al., 2021]{barrero2021working}
Barrero, J.~M., Bloom, N., and Davis, S.~J. (2021).
\newblock Why working from home will stick.
\newblock Technical report, National Bureau of Economic Research.

\bibitem[Bartosiewicz and Wi{\'s}niewski, 2015]{bartosiewicz2015use}
Bartosiewicz, B. and Wi{\'s}niewski, S. (2015).
\newblock The use of modern information technology in research on transport
  accessibility.
\newblock {\em Transport Problems}, 10.

\bibitem[Beck and Hensher, 2021]{Beck_Hensher_2021}
Beck, M.~J. and Hensher, D.~A. (2021).
\newblock {What might the changing incidence of Working from Home (WFH) tell us
  about future transport and land use agendas}.
\newblock {\em Transport Reviews}, 41(3):257–261.

\bibitem[Bernardino and Ben-Akiva, 1996]{Bernardino_Ben-Akiva_1996}
Bernardino, A. and Ben-Akiva, M. (1996).
\newblock Modeling the process of adoption of telecommuting: Comprehensive
  framework.
\newblock {\em Transportation Research Record}, 1552(1):161–170.

\bibitem[Bieser et~al., 2021]{bieser2021impacts}
Bieser, J.~C., Vaddadi, B., Kramers, A., H{\"o}jer, M., and Hilty, L.~M.
  (2021).
\newblock {Impacts of telecommuting on time use and travel: A case study of a
  neighborhood telecommuting center in Stockholm}.
\newblock {\em Travel Behaviour and Society}, 23:157--165.

\bibitem[Borkowski et~al., 2021]{borkowski2021lockdowned}
Borkowski, P., Ja{\.z}d{\.z}ewska-Gutta, M., and Szmelter-Jarosz, A. (2021).
\newblock {Lockdowned: Everyday mobility changes in response to COVID-19}.
\newblock {\em Journal of Transport Geography}, 90:102906.

\bibitem[{Boston University}, 2021]{bu2021}
{Boston University} (2021).
\newblock Future of staff work survey report.
\newblock
  \url{https://www.bu.edu/hr/documents/BU_Future_of_Staff_Work_Survey_Report_2021.pdf}.
\newblock Online; accessed on 2021-07-09.

\bibitem[Brewer, 1998]{Brewer_1998}
Brewer, A.~M. (1998).
\newblock Work design, flexible work arrangements and travel behaviour: Policy
  implications.
\newblock {\em Transport Policy}, 5(2):93–101.

\bibitem[Brewer and Hensher, 2002]{brewer2002flexible}
Brewer, A.~M. and Hensher, D.~A. (2002).
\newblock Flexible work and travel behaviour: A research framework.
\newblock In {\em Teleworking}, pages 235--252. Routledge.

\bibitem[Brock and Durlauf, 2001]{brock2001discrete}
Brock, W.~A. and Durlauf, S.~N. (2001).
\newblock Discrete choice with social interactions.
\newblock {\em The Review of Economic Studies}, 68(2):235--260.

\bibitem[Chavez-Dreyfuss, 2020]{chavez2020}
Chavez-Dreyfuss, G. (2020).
\newblock Permanently remote workers seen doubling in 2021 due to pandemic
  productivity: Survey.
\newblock {\em Reuters}.
\newblock Online; accessed on 2021-07-09.

\bibitem[Choo et~al., 2005]{choo2005does}
Choo, S., Mokhtarian, P.~L., and Salomon, I. (2005).
\newblock {Does telecommuting reduce vehicle-miles traveled? An aggregate time
  series analysis for the US}.
\newblock {\em Transportation}, 32(1):37--64.

\bibitem[Choudhury et~al., 2021]{choudhury2021work}
Choudhury, P., Foroughi, C., and Larson, B. (2021).
\newblock Work-from-anywhere: The productivity effects of geographic
  flexibility.
\newblock {\em Strategic Management Journal}, 42(4):655--683.

\bibitem[Choudhury, 2020]{Choudhury_2020}
Choudhury, P.~R. (2020).
\newblock Our work-from-anywhere future.
\newblock {\em Harvard Business Review}.

\bibitem[Cleary et~al., 2010]{cleary2010more}
Cleary, N., Worthington-Eyre, H., and Marinelli, P. (2010).
\newblock {More flex in the city: A case study from Brisbane of spreading the
  load in the office and on the road}.
\newblock In {\em Australasian Transport Research Forum (ATRF), 33rd, 2010,
  Canberra, ACT, Australia}.

\bibitem[Coenen and Kok, 2014]{Coenen_Kok_2014}
Coenen, M. and Kok, R. A.~W. (2014).
\newblock {Workplace flexibility and new product development performance: The
  role of telework and flexible work schedules}.
\newblock {\em European Management Journal}, 32(4):564–576.

\bibitem[Cutter, 2021]{cutterWSJ}
Cutter, C. (2021).
\newblock Facebook lets more employees choose full-time remote work or return
  to the office.
\newblock {\em Wall Street Journal}.

\bibitem[Dal~Fiore et~al., 2014]{DalFiore_Mokhtarian_Salomon_Singer_2014}
Dal~Fiore, F., Mokhtarian, P.~L., Salomon, I., and Singer, M.~E. (2014).
\newblock {“Nomads at last”? A set of perspectives on how mobile technology
  may affect travel}.
\newblock {\em Journal of Transport Geography}, 41:97–106.

\bibitem[de~Abreu~e Silva and Melo, 2018a]{deAbreuSilva_Melo_2018b}
de~Abreu~e Silva, J. and Melo, P.~C. (2018a).
\newblock {Does home-based telework reduce household total travel? A path
  analysis using single and two worker British households}.
\newblock {\em Journal of Transport Geography}, 73:148–162.

\bibitem[de~Abreu~e Silva and Melo, 2018b]{deAbreuSilva_Melo_2018}
de~Abreu~e Silva, J. and Melo, P.~C. (2018b).
\newblock Home telework, travel behavior, and land-use patterns: A path
  analysis of british single-worker households.
\newblock {\em Journal of Transport and Land Use}, 11(1):419–441.

\bibitem[De~Graaff and Rietveld, 2007]{de2007substitution}
De~Graaff, T. and Rietveld, P. (2007).
\newblock {Substitution between working at home and out-of-home: The role of
  ICT and commuting costs}.
\newblock {\em Transportation Research Part A: Policy and Practice},
  41(2):142--160.

\bibitem[Franch-Pardo et~al., 2020]{franch2020spatial}
Franch-Pardo, I., Napoletano, B.~M., Rosete-Verges, F., and Billa, L. (2020).
\newblock Spatial analysis and gis in the study of covid-19. a review.
\newblock {\em Science of The Total Environment}, 739:140033.

\bibitem[Galanti et~al., 2021]{galanti2021work}
Galanti, T., Guidetti, G., Mazzei, E., Zappal{\`a}, S., and Toscano, F. (2021).
\newblock Work from home during the covid-19 outbreak: The impact on
  employees’ remote work productivity, engagement, and stress.
\newblock {\em Journal of Occupational and Environmental Medicine}, 63(7):e426.

\bibitem[Gandini, 2015]{gandini2015rise}
Gandini, A. (2015).
\newblock The rise of coworking spaces: A literature review.
\newblock {\em Ephemera: Theory and Politics in Organization}, 15(1):193--205.

\bibitem[Ge et~al., 2018]{Ge_Polhill_Craig_2018}
Ge, J., Polhill, J.~G., and Craig, T.~P. (2018).
\newblock {Too much of a good thing? Using a spatial agent-based model to
  evaluate “unconventional” workplace sharing programmes}.
\newblock {\em Journal of Transport Geography}, 69:83–97.

\bibitem[George et~al., 2021]{george2021supporting}
George, T.~J., Atwater, L.~E., Maneethai, D., and Madera, J.~M. (2021).
\newblock Supporting the productivity and wellbeing of remote workers: Lessons
  from covid-19.
\newblock {\em Organizational Dynamics}, page 100869.

\bibitem[Girit, 2013]{Girit_2013}
Girit, D. (2013).
\newblock {\em A comparison of telecommuting and workplace employees in terms
  of situational strength, personality, work attitudes and performance}.
\newblock PhD thesis, Middle East Technical University.

\bibitem[Glaeser, 2010]{glaeser2010agglomeration}
Glaeser, E.~L. (2010).
\newblock {\em Agglomeration economics}.
\newblock University of Chicago Press.

\bibitem[Habib and Anik, 2021]{habib2021examining}
Habib, M.~A. and Anik, M. A.~H. (2021).
\newblock Examining the long term impacts of covid-19 using an integrated
  transport and land-use modelling system.
\newblock {\em International Journal of Urban Sciences}, pages 1--24.

\bibitem[Handy and Mokhtarian, 1995]{Handy_Mokhtarian_1995}
Handy, S.~L. and Mokhtarian, P.~L. (1995).
\newblock Planning for telecommuting measurement and policy issues.
\newblock {\em Journal of the American Planning Association}, 61(1):99–111.

\bibitem[Handy and Mokhtarian, 1996a]{Handy_Mokhtarian_1996b}
Handy, S.~L. and Mokhtarian, P.~L. (1996a).
\newblock Forecasting telecommuting: An exploration of methodologies and
  research needs.
\newblock {\em Transportation}, 23(2).

\bibitem[Handy and Mokhtarian, 1996b]{Handy_Mokhtarian_1996a}
Handy, S.~L. and Mokhtarian, P.~L. (1996b).
\newblock The future of telecommuting.
\newblock {\em Futures}, 28(3):227–240.

\bibitem[Harkness, 1977]{harkness1977selected}
Harkness, R.~C. (1977).
\newblock Selected results from a technology assessment of
  telecommunication-transportation interactions.
\newblock In {\em The Many Facets of Human Settlements}, pages 37--48.
  Elsevier.

\bibitem[{Harvard Business School Online}, 2021]{hbs2021}
{Harvard Business School Online} (2021).
\newblock {HBS Online} survey shows most professionals have excelled while
  working from home.
\newblock \url{https://online.hbs.edu/blog/post/future-of-work-from-home}.
\newblock Online; accessed on 2021-07-09.

\bibitem[Helling and Mokhtarian, 2001]{Helling_Mokhtarian_2001}
Helling, A. and Mokhtarian, P.~L. (2001).
\newblock Worker telecommunication and mobility in transition: Consequences for
  planning.
\newblock {\em Journal of Planning Literature}, 15(4):511–525.

\bibitem[Hong, 2002]{hong2002travel}
Hong, Q. (2002).
\newblock {\em The travel behavior of home-based teleworkers}.
\newblock PhD thesis, University of Michigan.

\bibitem[Illegems et~al., 2001]{Illegems_Verbeke_SJegers_2001}
Illegems, V., Verbeke, A., and S’Jegers, R. (2001).
\newblock The organizational context of teleworking implementation.
\newblock {\em Technological Forecasting and Social Change}, 68(3):275–291.

\bibitem[Julsrud et~al., 2012]{julsrud2012business}
Julsrud, T.~E., Hjorthol, R., and Denstadli, J.~M. (2012).
\newblock Business meetings: do new videoconferencing technologies change
  communication patterns?
\newblock {\em Journal of Transport Geography}, 24:396--403.

\bibitem[Kaleba, 2021]{michigan2021}
Kaleba, N. (2021).
\newblock Staff survey highlights interest in remote work, other concerns.
\newblock {\em The University Record}.
\newblock Online; accessed on 2021-07-10.

\bibitem[Kim et~al., 2015]{Kim_Choo_Mokhtarian_2015}
Kim, S.-N., Choo, S., and Mokhtarian, P.~L. (2015).
\newblock Home-based telecommuting and intra-household interactions in work and
  non-work travel: A seemingly unrelated censored regression approach.
\newblock {\em Transportation Research Part A: Policy and Practice},
  80:197–214.

\bibitem[Koh et~al., 2013]{Koh_Allen_Zafar_2013}
Koh, C.-W., Allen, T., and Zafar, N. (2013).
\newblock Dissecting reasons for not telecommuting: Are nonusers a homogenous
  group?
\newblock {\em The Psychologist-Manager Journal}, 16:243–260.

\bibitem[Kunesh, 2021]{kunesh2021}
Kunesh, A. (2021).
\newblock You can now get \$50 monthly uber cash, uber pass subscription with
  wework all access.
\newblock {\em MSN Money}.

\bibitem[Laumer and Maier, 2021]{Laumer_Maier_2021}
Laumer, S. and Maier, C. (2021).
\newblock {Why do people (not) want to work from home? An individual-focused
  literature review on telework}.
\newblock In {\em Proceedings of the 2021 on Computers and People Research
  Conference}, page 41–49. ACM.

\bibitem[Litman, 2021]{litman2015evaluating}
Litman, T. (2021).
\newblock {\em Evaluating public transit benefits and costs}.
\newblock Victoria Transport Policy Institute, Victoria, BC, Canada.

\bibitem[Lund and Mokhtarian, 1994]{Lund_Mokhtarian_1994}
Lund, J.~R. and Mokhtarian, P.~L. (1994).
\newblock Telecommuting and residential location: Theory and implications for
  commute travel in monocentric metropolis.
\newblock {\em Transportation Research Record}, 1463.

\bibitem[Mariotti et~al., 2021a]{mariotti2021geography}
Mariotti, I., Akhavan, M., and Di~Matteo, D. (2021a).
\newblock The geography of coworking spaces and the effects on the urban
  context: Are pole areas gaining?
\newblock {\em New Workplaces—Location Patterns, Urban Effects and
  Development Trajectories: A Worldwide Investigation}, pages 169--194.

\bibitem[Mariotti et~al., 2021b]{mariotti2021research}
Mariotti, I., Akhavan, M., and Di~Vita, S. (2021b).
\newblock A research agenda for the future of workplaces.
\newblock {\em New Workplaces—Location Patterns, Urban Effects and
  Development Trajectories: A Worldwide Investigation}, pages 299--304.

\bibitem[Martin and MacDonnell, 2012]{martin2012telework}
Martin, B.~H. and MacDonnell, R. (2012).
\newblock Is telework effective for organizations?
\newblock {\em Management Research Review}.

\bibitem[Mokhtarian, 1990]{mokhtarian1990typology}
Mokhtarian, P.~L. (1990).
\newblock A typology of relationships between telecommunications and
  transportation.
\newblock {\em Transportation Research Part A: General}, 24(3):231--242.

\bibitem[Mokhtarian, 1991a]{mokhtarian1991defining}
Mokhtarian, P.~L. (1991a).
\newblock Defining telecommuting.
\newblock {\em Transportation Research Record}, 1305:273--281.

\bibitem[Mokhtarian, 1991b]{mokhtarian1991empirical}
Mokhtarian, P.~L. (1991b).
\newblock An empirical analysis of the transportation impacts of telecommuting.
\newblock Technical Report UCTC No. 131, University of California
  Transportation Research Center.
\newblock Working Paper.

\bibitem[Mokhtarian, 1991c]{mokhtarian1991telecommuting}
Mokhtarian, P.~L. (1991c).
\newblock Telecommuting and travel: State of the practice, state of the art.
\newblock {\em Transportation}, 18(4):319--342.

\bibitem[Mokhtarian, 1998]{mokhtarian1998synthetic}
Mokhtarian, P.~L. (1998).
\newblock A synthetic approach to estimating the impacts of telecommuting on
  travel.
\newblock {\em Urban Studies}, 35(2):215--241.

\bibitem[Mokhtarian, 2002]{mokhtarian2002telecommunications}
Mokhtarian, P.~L. (2002).
\newblock Telecommunications and travel: The case for complementarity.
\newblock {\em Journal of Industrial Ecology}, 6(2):43--57.

\bibitem[Mokhtarian and Bagley, 2000]{Mokhtarian_Bagley_2000}
Mokhtarian, P.~L. and Bagley, M.~N. (2000).
\newblock Modeling employees’ perceptions and proportional preferences of
  work locations: The regular workplace and telecommuting alternatives.
\newblock {\em Transportation Research Part A: Policy and Practice},
  34(4):223–242.

\bibitem[Mokhtarian et~al., 1995]{mokhtarian1995methodological}
Mokhtarian, P.~L., Handy, S.~L., and Salomon, I. (1995).
\newblock Methodological issues in the estimation of the travel, energy, and
  air quality impacts of telecommuting.
\newblock {\em Transportation Research Part A: Policy and Practice},
  29(4):283--302.

\bibitem[Mokhtarian and Salomon, 1994]{Mokhtarian_Salomon_1994}
Mokhtarian, P.~L. and Salomon, I. (1994).
\newblock Modeling the choice of telecommuting: Setting the context.
\newblock {\em Environment and Planning A: Economy and Space}, 26(5):749–766.

\bibitem[Mokhtarian and Salomon, 1996a]{Mokhtarian_Salomon_1996a}
Mokhtarian, P.~L. and Salomon, I. (1996a).
\newblock {Modeling the choice of telecommuting: 2. A case of the preferred
  impossible alternative}.
\newblock {\em Environment and Planning A}, 28(10):1859–1876.

\bibitem[Mokhtarian and Salomon, 1996b]{Mokhtarian_Salomon_1996b}
Mokhtarian, P.~L. and Salomon, I. (1996b).
\newblock {Modeling the choice of telecommuting: 3. Identifying the choice set
  and estimating binary choice models for technology-based alternatives}.
\newblock {\em Environment and Planning A: Economy and Space},
  28(10):1877–1894.

\bibitem[Mokhtarian and Salomon, 1997a]{Mokhtarian_Salomon_1997}
Mokhtarian, P.~L. and Salomon, I. (1997a).
\newblock Modeling the desire to telecommute: The importance of attitudinal
  factors in behavioral models.
\newblock {\em Transportation Research Part A: Policy and Practice},
  31(1):35–50.

\bibitem[Mokhtarian and Salomon, 1997b]{mokhtarian1997modeling}
Mokhtarian, P.~L. and Salomon, I. (1997b).
\newblock Modeling the desire to telecommute: The importance of attitudinal
  factors in behavioral models.
\newblock {\em Transportation Research Part A: Policy and Practice},
  31(1):35--50.

\bibitem[Mokhtarian et~al., 2005]{Mokhtarian_Salomon_Choo_2005}
Mokhtarian, P.~L., Salomon, I., and Choo, S. (2005).
\newblock {Measuring the measurable: Why can’t we agree on the number of
  telecommuters in the U.S.?}
\newblock {\em Quality \& Quantity}, 39(4):423–452.

\bibitem[Mokhtarian and Varma, 1998]{Mokhtarian_Varma_1998}
Mokhtarian, P.~L. and Varma, K.~V. (1998).
\newblock The trade-off between trips and distance traveled in analyzing the
  emissions impacts of center-based telecommuting.
\newblock {\em Transportation Research Part D: Transport and Environment},
  3(6):419–428.

\bibitem[Muhammad et~al., 2008]{muhammad2008modelling}
Muhammad, S., de~Jong, T., and Ottens, H.~F. (2008).
\newblock Job accessibility under the influence of information and
  communication technologies, in the netherlands.
\newblock {\em Journal of Transport Geography}, 16(3):203--216.

\bibitem[Nagurney et~al., 2002]{Nagurney_Dong_Mokhtarian_2002}
Nagurney, A., Dong, J., and Mokhtarian, P.~L. (2002).
\newblock Multicriteria network equilibrium modeling with variable weights for
  decision-making in the information age with applications to telecommuting and
  teleshopping.
\newblock {\em Journal of Economic Dynamics and Control}, 26(9):1629–1650.

\bibitem[Nagurney et~al., 2003]{Nagurney_Dong_Mokhtarian_2003}
Nagurney, A., Dong, J., and Mokhtarian, P.~L. (2003).
\newblock A space-time network for telecommuting versus commuting
  decision-making.
\newblock {\em Papers in Regional Science}, 82(4):451–473.

\bibitem[Nayak and Pandit, 2021]{Nayak_Pandit_2021}
Nayak, S. and Pandit, D. (2021).
\newblock {Potential of telecommuting for different employees in the Indian
  context beyond COVID-19 lockdown}.
\newblock {\em Transport Policy}.

\bibitem[Nilles, 1988]{Nilles_1988}
Nilles, J.~M. (1988).
\newblock {Traffic reduction by telecommuting: A status review and selected
  bibliography}.
\newblock {\em Transportation Research Part A: General}, 22(4):301–317.

\bibitem[Oliker and Bekhor, 2018]{oliker2018frequency}
Oliker, N. and Bekhor, S. (2018).
\newblock A frequency based transit assignment model that considers online
  information.
\newblock {\em Transportation Research Part C: Emerging Technologies},
  88:17--30.

\bibitem[Olson, 1983]{Olson_1983}
Olson, M.~H. (1983).
\newblock Remote office work: Changing work patterns in space and time.
\newblock {\em Communications of the ACM}, 26(3):182–187.

\bibitem[Oppenheim, 1993]{oppenheim1993equilibrium}
Oppenheim, N. (1993).
\newblock Equilibrium trip distribution/assignment with variable destination
  costs.
\newblock {\em Transportation Research Part B: Methodological}, 27(3):207--217.

\bibitem[Ory and Mokhtarian, 2006]{Ory_Mokhtarian_2006}
Ory, D.~T. and Mokhtarian, P.~L. (2006).
\newblock {Which came first, the telecommuting or the residential relocation?
  An empirical analysis of causality}.
\newblock {\em Urban Geography}, 27(7):590–609.

\bibitem[Ozimek, 2020]{ozimek2020future}
Ozimek, A. (2020).
\newblock The future of remote work.
\newblock {\em Available at SSRN 3638597}.

\bibitem[Paleti, 2016]{paleti2016generalized}
Paleti, R. (2016).
\newblock Generalized extreme value models for count data: Application to
  worker telecommuting frequency choices.
\newblock {\em Transportation Research Part B: Methodological}, 83:104--120.

\bibitem[Pawlak et~al., 2015]{Pawlak_Polak_Sivakumar_2015}
Pawlak, J., Polak, J.~W., and Sivakumar, A. (2015).
\newblock {Towards a microeconomic framework for modelling the joint choice of
  activity–travel behaviour and ICT use}.
\newblock {\em Transportation Research Part A: Policy and Practice},
  76:92–112.

\bibitem[Plyushteva, 2019]{plyushteva2019commutes}
Plyushteva, A. (2019).
\newblock Commutes and co-workers: Complicating individual journeys through
  workplace relations.
\newblock {\em Built Environment}, 45(4):603--620.

\bibitem[Pouri and Bhat, 2003]{Pouri_Bhat_2003}
Pouri, Y.~D. and Bhat, C.~R. (2003).
\newblock On modeling choice and frequency of home-based telecommuting.
\newblock {\em Transportation Research Record}, 1858(1):55–60.

\bibitem[Pratt, 2003]{pratt2003telework}
Pratt, J.~H. (2003).
\newblock {Telework trends in the United States}.
\newblock In {\em Organisation and Work Beyond 2000}, pages 345--356. Springer.

\bibitem[{PwC US}, 2021]{pwc2021}
{PwC US} (2021).
\newblock {PwC US Remote Work Survey}.
\newblock \url{https://www.pwc.com/us/remotework}.
\newblock Online; accessed on 2021-07-09.

\bibitem[Raghuram et~al., 2018]{Raghuram_Hill_Gibbs_Maruping_2018}
Raghuram, S., Hill, N., Gibbs, J., and Maruping, L. (2018).
\newblock Virtual work: Bridging research clusters.
\newblock {\em Academy of Management Annals}, 13.

\bibitem[Rosenthal et~al., 2021]{rosenthal2021jue}
Rosenthal, S.~S., Strange, W.~C., and Urrego, J.~A. (2021).
\newblock Jue insight: Are city centers losing their appeal? commercial real
  estate, urban spatial structure, and covid-19.
\newblock {\em Journal of Urban Economics}, page 103381.

\bibitem[Ross and Ressia, 2015]{Ross_Ressia_2015}
Ross, P. and Ressia, S. (2015).
\newblock Neither office nor home: Coworking as an emerging workplace choice.
\newblock {\em Employment Relations Record}, 15(1):42--57.

\bibitem[Salomon and Mokhtarian, 2008]{Salomon_Mokhtarian_2008}
Salomon, I. and Mokhtarian, P. (2008).
\newblock {Can telecommunications help solve transportation problems? A decade
  later: are the prospects any better?}
\newblock {\em Handbook of Transport Modelling}.

\bibitem[Saxena and Mokhtarian, 1997]{Saxena_Mokhtarian_1997}
Saxena, S. and Mokhtarian, P.~L. (1997).
\newblock The impact of telecommuting on the activity spaces of participants.
\newblock {\em Geographical Analysis}, 29(2):124–144.

\bibitem[Schramm et~al., 2021]{duke2021}
Schramm, S., Black, J., and Minai, L. (2021).
\newblock A remote work future at {Duke}.
\newblock {\em Duke Today}.
\newblock Online; accessed on 2021-07-09.

\bibitem[Sener and Bhat, 2011]{Sener_Bhat_2011}
Sener, I.~N. and Bhat, C.~R. (2011).
\newblock A copula-based sample selection model of telecommuting choice and
  frequency.
\newblock {\em Environment and Planning A}, 43(1):126–145.

\bibitem[Shafizadeh et~al., 2007]{Shafizadeh_Niemeier_Mokhtarian_Salomon_2007}
Shafizadeh, K.~R., Niemeier, D.~A., Mokhtarian, P.~L., and Salomon, I. (2007).
\newblock {Costs and benefits of home-based telecommuting: A Monte Carlo
  simulation model incorporating telecommuter, employer, and public sector
  perspectives}.
\newblock {\em Journal of Infrastructure Systems}, 13(1):12–25.

\bibitem[Shamir and Salomon, 1985]{shamir1985work}
Shamir, B. and Salomon, I. (1985).
\newblock Work-at-home and the quality of working life.
\newblock {\em Academy of Management Review}, 10(3):455--464.

\bibitem[Shearmur et~al., 2021]{shearmur2021towards}
Shearmur, R., Ananian, P., Lachapelle, U., Parra-Lokhorst, M., Paulhiac, F.,
  Tremblay, D.-G., and Wycliffe-Jones, A. (2021).
\newblock Towards a post-covid geography of economic activity: Using
  probability spaces to decipher montreal’s changing workscapes.
\newblock {\em Urban Studies}, page 00420980211022895.

\bibitem[Singh et~al., 2013]{Singh_Paleti_Jenkins_Bhat_2013}
Singh, P., Paleti, R., Jenkins, S., and Bhat, C.~R. (2013).
\newblock On modeling telecommuting behavior: Option, choice, and frequency.
\newblock {\em Transportation}, 40(2):373–396.

\bibitem[Smith, 2016]{smith2016gig}
Smith, A. (2016).
\newblock Gig work, online selling and home sharing.
\newblock Technical report, Pew Research Center.

\bibitem[Stanek and Mokhtarian, 1998]{stanek1998developing}
Stanek, D.~M. and Mokhtarian, P.~L. (1998).
\newblock Developing models of preference for home-based and center-based
  telecommuting: Findings and forecasts.
\newblock {\em Technological Forecasting and Social Change}, 57(1-2):53--74.

\bibitem[Stiles, 2019]{stiles2019working}
Stiles, J. (2019).
\newblock {\em Working at home and elsewhere in the city: mobile cloud
  computing, telework, and urban travel}.
\newblock PhD thesis, Rutgers University School of Graduate Studies.

\bibitem[Stiles and Smart, 2020]{Stiles_Smart_2020}
Stiles, J. and Smart, M.~J. (2020).
\newblock {Working at home and elsewhere: Daily work location, telework, and
  travel among United States knowledge workers}.
\newblock {\em Transportation}.

\bibitem[Strack et~al., 2021]{bcg2020}
Strack, R., Kovács-Ondrejkovic, R., Baier, J., Antebi, P., Kavanagh, K., and
  Gobernado, A.~L. (2021).
\newblock Decoding global ways of working.
\newblock
  \url{https://www.bcg.com/en-us/publications/2021/advantages-of-remote-work-flexibility}.
\newblock Online; accessed on 2021-07-09.

\bibitem[Su et~al., 2021]{Su_McBride_Goulias_2021}
Su, R., McBride, E.~C., and Goulias, K.~G. (2021).
\newblock Unveiling daily activity pattern differences between telecommuters
  and commuters using human mobility motifs and sequence analysis.
\newblock {\em Transportation Research Part A: Policy and Practice},
  147:106–132.

\bibitem[Tang et~al., 2011]{Tang_Mokhtarian_Handy_2011}
Tang, W.~L., Mokhtarian, P.~L., and Handy, S.~L. (2011).
\newblock {The impact of the residential built environment on work at home
  adoption and frequency: An example from Northern California}.
\newblock {\em Journal of Transport and Land Use}, 4(3):3–22.

\bibitem[{The Economist}, 2021]{economist2021}
{The Economist} (2021).
\newblock Office re-entry is proving trickier than last year’s abrupt exit.
\newblock {\em The Economist}.
\newblock Online; accessed on 2021-07-10.

\bibitem[Vakilian and Edrisi, 2019]{vakilian2019modeling}
Vakilian, R. and Edrisi, A. (2019).
\newblock Modeling factors affecting the choice of telework and its impact on
  demand in transportation networks.
\newblock {\em Revista Innovaciencia}, 7(2).

\bibitem[van Wee et~al., 2012]{van2012ict}
van Wee, B., Chorus, C., and Geurs, K.~T. (2012).
\newblock Ict and accessibility: research synthesis and future perspectives.
\newblock In {\em Accessibility analysis and transport planning}. Edward Elgar
  Publishing.

\bibitem[van Wee et~al., 2013]{van2013information}
van Wee, B., Geurs, K., and Chorus, C. (2013).
\newblock Information, communication, travel behavior and accessibility.
\newblock {\em Journal of Transport and Land Use}, 6(3):1--16.

\bibitem[van Wee and Witlox, 2021]{vanWee_Witlox_2021}
van Wee, B. and Witlox, F. (2021).
\newblock Covid-19 and its long-term effects on activity participation and
  travel behaviour: A multiperspective view.
\newblock {\em Journal of Transport Geography}, page 103144.

\bibitem[Vana et~al., 2008]{Vana_Bhat_Mokhtarian_2008}
Vana, P., Bhat, C.~R., and Mokhtarian, P.~L. (2008).
\newblock On modeling the choices of work-hour arrangement, location, and
  frequency of telecommuting.
\newblock In {\em 87th Annual Meeting of the Transport Research Board (TRB),
  Washington, DC}.

\bibitem[Wiener, 1950]{wiener1950speech}
Wiener, N. (1950).
\newblock Speech, language, and learning.
\newblock {\em The Journal of the Acoustical Society of America},
  22(6):696--697.

\bibitem[Yang et~al., 2021]{yang2021effects}
Yang, L., Holtz, D., Jaffe, S., Suri, S., Sinha, S., Weston, J., Joyce, C.,
  Shah, N., Sherman, K., Hecht, B., et~al. (2021).
\newblock The effects of remote work on collaboration among information
  workers.
\newblock {\em Nature human behaviour}, pages 1--12.

\bibitem[Yen et~al., 1994]{yen1994employer}
Yen, J.-R., Mahmassani, H.~S., and Herman, R. (1994).
\newblock Employer attitudes and stated preferences toward telecommuting: An
  exploratory analysis.
\newblock {\em Transportation Research Record}, 1463:15.

\bibitem[Yeraguntla and Bhat, 2005]{yeraguntla2005classification}
Yeraguntla, A. and Bhat, C.~R. (2005).
\newblock Classification taxonomy and empirical analysis of work arrangements.
\newblock {\em Transportation Research Record}, 1926(1):233--241.

\bibitem[Yu et~al., 2019]{Yu_Burke_Raad_2019}
Yu, R., Burke, M., and Raad, N. (2019).
\newblock Exploring impact of future flexible working model evolution on urban
  environment, economy and planning.
\newblock {\em Journal of Urban Management}, 8(3):447–457.

\bibitem[Zemo and Termansen, 2018]{zemo2018farmers}
Zemo, K.~H. and Termansen, M. (2018).
\newblock Farmers’ willingness to participate in collective biogas
  investment: A discrete choice experiment study.
\newblock {\em Resource and Energy Economics}, 52:87--101.

\bibitem[Zhang et~al., 2009]{zhang2009modeling}
Zhang, J., Kuwano, M., Lee, B., and Fujiwara, A. (2009).
\newblock Modeling household discrete choice behavior incorporating
  heterogeneous group decision-making mechanisms.
\newblock {\em Transportation Research Part B: Methodological}, 43(2):230--250.

\bibitem[Zhang and Zhang, 2021]{zhang2021long}
Zhang, R. and Zhang, J. (2021).
\newblock {Long-term pathways to deep decarbonization of the transport sector
  in the post-COVID world}.
\newblock {\em Transport Policy}.

\bibitem[Zhu et~al., 2018]{zhu2018metropolitan}
Zhu, P., Wang, L., Jiang, Y., and Zhou, J. (2018).
\newblock Metropolitan size and the impacts of telecommuting on personal
  travel.
\newblock {\em Transportation}, 45(2):385--414.

\end{thebibliography}
\end{document}